\documentclass[prd,onecolumn,showpacs,nofootinbib]{revtex4-1}
\usepackage{amsmath}
\usepackage{latexsym}
\usepackage{amsfonts}
\usepackage{graphicx}
\usepackage{mathrsfs}
\usepackage{hyperref}
\usepackage{mathrsfs}
\usepackage{accents}
\usepackage{color}

\begin{document}

\title{Gravitational waveforms from the quasicircular inspiral of compact binaries in massive Brans-Dicke theory}

\author{Tan Liu$^1$}\email{lewton@hust.edu.cn}
\author{Wen Zhao$^{2,3}$}
\author{Yan Wang$^1$}

\affiliation{$^1$MOE Key Laboratory of Fundamental Physical Quantities Measurements, Hubei Key
Laboratory of Gravitation and Quantum Physics, PGMF and School of Physics, Huazhong
University of Science and Technology, Wuhan 430074, China}
\affiliation{$^2$CAS Key Laboratory for Researches in Galaxies and Cosmology, Department of Astronomy, \\
University of Science and Technology of China, Chinese Academy of Sciences, Hefei, Anhui 230026, China}
\affiliation{$^3$School of Astronomy and Space Science, University of Science and Technology of China, Hefei 230026, China}

\begin{abstract}
We study the gravitational waves emitted by an inspiralling compact binary system in massive Brans-Dicke theory. In addition to the two tensor polarizations, which have been obtained in the previous work, we calculate explicitly and analytically the expressions for the time-domain waveforms of the two scalar polarizations.  With the stationary phase approximations, we obtain the Fourier transforms of the two tensor polarizations. We find that when the scalar field is light, the waveforms can be mapped to the parametrized post-Einsteinian (ppE) framework and we identify the ppE parameters. However, when the scalar field is heavy, the ppE framework is not applicable. We also obtain the projected constraints on the parameters of this theory by gravitational wave observations of  future ground-based detectors. Finally, we apply our result to the model proposed by Damour and Esposito-Far\`{e}se, $f(R)$ gravity, and screened modified gravity.
\end{abstract}

\pacs{98.70.Vc, 98.80.Cq, 04.30.-w}

\maketitle

\section{Introduction}
The direct detection by the LIGO-Virgo collaboration of the gravitational waves (GWs) emitted by a binary black hole system has opened a new window to test gravity in the strong and dynamical regime \cite{PhysRevLett.116.061102}. Although general relativity (GR), as the most successful theory of gravity, has past all the observational constraints, it still has a lot of shortcomings \cite{Berti_2015}. 
Based on  different theoretical (e.g. a quantum theory of gravity) and observational (e.g. the accelerating expansion of the universe) considerations, various extensions of GR have been proposed \cite{Barack_2019,CLIFTON20121}.

In this work, we focus on the gravitational waves emitted by a binary system in  massive Brans-Dicke theory, an extension of GR with a massive scalar field \cite{Bergmann1968}. In GR, GW has only two tensor polarizations  ($h_+$ and $h_\times$) and gravitational radiation begins at the quadrupole order \cite{misner1973gravitation}. While in massive Brans-Dicke theory,  the scalar field can introduce extra GW polarizations and dipole radiation.  Thus far, a number of studies have investigated the effects of the scalar field on the motion and gravitational radiation of a binary system \cite{Damour_1992}. When the velocity of the binary system is slow and the gravitational field is weak, post-Newtonian (PN) expansion method can be used to compute the orbital motion and gravitational radiation \cite{Poisson2014,Levi_2020}. The time derivative of the  orbit period of the binary system due to  the scalar dipole radiation was worked out by Eardley \cite{1975ApJ...196L..59E}. The GW waveforms emitted by a binary system in Brans-Dicke theory have been calculated to the Newtonian quadrupole order in \cite{PhysRevD.50.6058}. Then, by adapting the direct integration of the relaxed Einstein equations formalism to Brans-Dicke theory, the scalar waveform was calculated to 1.5PN order and the tensor waveform was calculated to 2PN order \cite{Will-BDn,PhysRevD.89.084014,Lang-BD,PhysRevD.94.084003}. Using the Fokker action of point particles, the equation of motion of a binary system was obtained up to 3PN order \cite{PhysRevD.98.044004,PhysRevD.99.044047}. Recently, the tidal effect due to the scalar field, which starts at 3PN order, has been incorporated into the phase of the waveforms \cite{Bernard2020}.   

All the above works \cite{1975ApJ...196L..59E,PhysRevD.50.6058,Will-BDn,PhysRevD.89.084014,Lang-BD,PhysRevD.94.084003,PhysRevD.98.044004,PhysRevD.99.044047,Bernard2020} focused on the massless scalar field. For a massive scalar field, there exist some unique features. A massive scalar field can induce two polarizations, the breathing polarization $h_b$ and the longitudinal polarization $h_L$, while a massless scalar field induces only $h_b$ \cite{PhysRevD.62.024004}. The screen mechanism can be imposed when the scalar field is massive \cite{Burrage2018}. The scalar field can develop an environment dependent mass. In the high density environment, the scalar field field is heavy and the scalar force is screened. It can help to pass the local solar system test. In the low density cosmological background, the scalar field is light and the scalar force is long range which can accelerate the universe. A scalar-tensor theory with the screen mechanism is called screened modified gravity (SMG). The four GW polarizations emitted by a binary system in the  screened modified gravity has been calculated to  the Newtonian quadrupole order \cite{Zhang2017,PhysRevD.98.083023,Niu_2020}. There is another surprising effect of the massive scalar field.
In an extreme mass ratio inspiral (EMRI) system, where a stellar mass object spirals into a supermassive black hole, the massive scalar field may produce floating orbits \cite{PhysRevLett.107.241101,PhysRevD.95.044016,PhysRevD.99.064018,PhysRevD.101.043020}. Due to superradiance, the energy flux emitted by the stellar mass object at the horizon of the rotating supermassive black hole may be negative. The negative energy flux at the horizon can compensate for the positive energy flux at infinity. Therefore, the orbital decay rate of the stellar mass object becomes zero and it floats around the supermassive black hole. For the EMRI on a quasicircular orbit,  the phase of the tensor waveforms has been  worked out by  the black hole perturbation method \cite{PhysRevD.85.102003}.

More recent attention has focused on the effects of the massive scalar field. For an EMRI system, the self-force equation for the stellar mass object moving on an accelerated world line in the black hole background spacetime has been obtained through the perturbation method \cite{PhysRevD.87.104020,PhysRevD.92.064051}.
The response of the gravitational wave interferometer to the massive scalar wave has been analyzed in \cite{PhysRevD.92.063013}. The gravitational radiation power and the tensor waveforms of a binary system in massive Brans-Dicke theory have been calculated to Newtonian quadrupole order \cite{PhysRevD.85.064041,PhysRevD.85.122005}. The effective field theory approach has been used to study this problem in \cite{PhysRevD.99.063013}. Using relativistic hydrodynamical simulations, the binary neutron star mergers in the presence of a massive scalar field have been studied numerically in \cite{PhysRevD.97.064016}

In this paper, we continue these efforts to study GWs in massive Brans-Dicke theory. We work out the GW waveforms emitted by an inspiral compact binary system on a quasicircular orbit in massive Brans-Dicke theory. We obtain the expressions of the four polarizations in the time domain.  The waveforms of the two tensor polarizations have been obtained in \cite{PhysRevD.85.122005}. The waveforms of the two scalar polarizations are the new results.
We  find that when the scalar field is light, the Fourier transforms of the tensor polarizations can be mapped to the parametrized post-Einsteinian (ppE) framework \cite{PhysRevD.80.122003}. We identify the ppE parameters in this situation. When the scalar field is heavy, the waveforms become complicated and the ppE framework is not applicable. We also study the constraints on the parameters of massive Brans-Dicke theory that future ground-based GW detectors will  impose. Then we apply our result to the model proposed by Damour and Esposito-Far\`{e}se \cite{PhysRevLett.70.2220} and its extension with a massive scalar field \cite{PhysRevD.93.064005,PhysRevD.96.084026}. Since $f(R)$ gravity can be rewritten as massive Brans-Dicke theory with coupling function $\omega(\phi)=0$ \cite{RevModPhys.82.451,de2010f,LIU2018286}, we also apply the result to $f(R)$ gravity. At last, we compare massive Brans-Dicke theory with SMG models,  including chameleon model \cite{PhysRevLett.93.171104,PhysRevD.69.044026} and symmetron model \cite{PhysRevLett.104.231301}.

The paper is organized as follows. Sections \ref{mBD} and \ref{eom} review the relevant results of \cite{PhysRevD.85.064041}. In Section \ref{mBD}, we rederive the weak-field expansion of the field equations.
In Section \ref{eom}, we investigate the motion of point particles.  Section \ref{gw_waveform} begins to demonstrate the new results.  In Section \ref{gw_waveform}, we obtain the GW waveforms of an inspiral compact binary.  In Section \ref{ppeob}, we compare our results with  the ppE framework and apply them to different models. Section \ref{con} concludes and points possible directions for future research. 

For the metric, Riemann and Ricci tensors, we follow the conventions of Misner, Thorne and Wheeler \cite{misner1973gravitation}. We set the units so that $c=\hbar=1$. We do not set  $G$ equal to 1, since the effective gravitational constant depends on the background value of the scalar field, which will vary over the history of the universe.

\section{ massive Brans-Dicke theory }\label{mBD}
In this section, we review some relevant results from \cite{PhysRevD.85.064041}. The action of massive Brans-Dicke theory in the Jordan frame takes the form \cite{PhysRevD.85.064041}
\begin{equation}\label{action}
S=\frac{1}{16\pi}\int d^4 x\sqrt{-g}\left[\phi R-\frac{\omega(\phi)}{\phi} \partial_\mu\phi\partial^\mu\phi+M(\phi)\right]+S_m\left[g_{\mu\nu},\Psi_m\right],
\end{equation}
where $g\equiv \det g_{\mu\nu}$ and   $\omega(\phi)$ is the coupling function which is
responsible for the spontaneous scalarization phenomenon \cite{PhysRevLett.70.2220}. The  function  $M(\phi)$ can provide the effective cosmological constant and the mass of the scalar field.  $\Psi_m$ denotes the matter fields collectively. The matter action for a system of pointlike particles can be written as \cite{weinberg1972gravitation}

\begin{equation}
S_m=-\sum_A \int m_A(\phi)\ d\tau_A
\end{equation}
where $\tau_A$ is the proper time of body A and the mass of body A depends on the scalar field $\phi$, because the scalar field can influence the self-gravity of the
compact object. This approach was first proposed by Eardley \cite{1975ApJ...196L..59E}. Gralla reproduced this relation in a more general framework \cite{PhysRevD.87.104020}. 
Variation of the action \eqref{action} yields the field equations \cite{PhysRevD.85.064041}

\begin{equation}\label{teq}
R_{\mu\nu}-\frac12 g_{\mu\nu}R-\frac12 \frac{M(\phi)}{\phi}g_{\mu\nu}=\frac{8\pi}{\phi}T_{\mu\nu}+\frac{\omega(\phi)}{\phi^2}(\phi_{,\mu}\phi_{,\nu}-\frac12 g_{\mu\nu}\phi_{,\alpha}\phi^{,\alpha})+\frac{1}{\phi}(\nabla_\mu\nabla_\nu-g_{\mu\nu} \square)\phi
\end{equation}

\begin{equation}\label{seq}
\square\phi+\frac{1}{2\omega(\phi)+3}(\phi M'-2M)=\frac{8\pi}{2\omega(\phi)+3}(T-2\phi~ T')-\frac{\omega'}{2\omega(\phi)+3}\phi_{,\alpha}\phi^{,\alpha}
\end{equation}
where $'\equiv\frac{d}{d\phi}$ and $\square\equiv\nabla_\nu\nabla^\nu$. The stress-energy tensor takes the form \cite{weinberg1972gravitation}
\begin{equation}\label{set}
T^{\mu\nu}=\frac{1}{\sqrt{-g}}\sum_A\frac{u_A^\mu u_A^\nu}{u_A^0}m_A(\phi)\delta^{(3)}({\bf x-x}_A),
\end{equation}
where $u_A^\mu$ is the four velocity of body A and ${\bf x}_A$ is its position. $T=g^{\mu\nu}T_{\mu\nu}$ is the trace of the stress-energy tensor.

Then we expand the the metric tensor $g_{\mu\nu}$ about the Minkowski background $\eta_{\mu\nu}$ and the scalar field $\phi$ around its constant background value $\phi_0$
\begin{equation}
\phi= \phi_0+\varphi, \qquad g_{\mu\nu}=\eta_{\mu\nu}+h_{\mu\nu},
\end{equation}
\begin{equation}\label{thetamunu}
\theta_{\mu\nu}\equiv h_{\mu\nu}-\frac12 h \eta_{\mu\nu}-\frac{\varphi}{\phi_0}\eta_{\mu\nu},
\end{equation}
where $\varphi$ , $h_{\mu\nu}$, and $\theta_{\mu\nu}$ are small perturbations. $h=\eta^{\mu\nu}h_{\mu\nu}$ is the trace of the metric perturbation. In terms of $\theta_{\mu\nu}$, the tensor field equation \eqref{teq}  can be transformed into  a standard wave equation. (For more details, see appendix C of \cite{Saffer_2018}.) 
In order to expand the field equations in the weak-field limit, we  need to expand the two functions $M(\phi)$ and $\omega(\phi)$ around the scalar background $\phi_0$
\begin{equation}
M(\phi)=M(\phi_0)+M'(\phi_0)\varphi+\frac12M''(\phi_0)\varphi^2+\cdots
\end{equation}
\begin{equation}
\omega(\phi)=\omega_0+\omega_1 \varphi+\cdots
\end{equation}
where $\omega_0\equiv\omega(\phi_0)$ and $\omega_1\equiv\omega'(\phi_0)$.

We assume that $g_{\mu\nu}=\eta_{\mu\nu}$ and $\phi=\phi_0$ is a vacuum solution of the field equations \eqref{teq} and \eqref{seq}. That is, the spacetime is asymptotically flat \cite{PhysRevLett.108.081103}. Therefore, we have \cite{PhysRevD.85.064041}
\begin{equation}
M(\phi_0)=M'(\phi_0)=0.
\end{equation}
We also need to expand the mass of point particle around the scalar background $\phi_0$
\begin{equation}\label{maphi}
m_A(\phi)=m_A\left[1+s_A\frac{\varphi}{\phi_0}+\frac12(s_A^2+s_A'-s_A)\big(\frac{\varphi}{\phi_0}\big)^2+\cdots\right]
\end{equation}
where $m_A\equiv m_A(\phi_0)$. $s_A$ and  $s_A'$ are the sensitivity and its derivative of point particle A \cite{1975ApJ...196L..59E},
\begin{equation}
s_A=\frac{d \ln m_A(\phi)}{d \ln \phi}\Big|_{\phi_0}, \qquad s_A'=\frac{d^2 \ln m_A(\phi)}{d(\ln \phi)^2}\Big|_{\phi_0} .
\end{equation}
The sensitivity of a black hole is $\frac12$  \cite{1975ApJ...196L..59E}. The typical value of the sensitivity of a neutron star is about 0.2 \cite{PhysRevD.85.064041}.

The tensor field equation \eqref{teq} in the weak-field limit becomes \cite{PhysRevD.85.064041}
\begin{equation}\label{weak-t}
\square_\eta \theta_{\mu\nu}=-16\pi \tau_{\mu\nu},
\end{equation}
where $\square_\eta=\eta^{\mu\nu}\partial_\mu\partial_\nu$ and  $\tau_{\mu\nu}=T_{\mu\nu}/\phi_0+t_{\mu\nu}$. $t_{\mu\nu}\equiv O(\theta^2,\varphi^2,\theta\varphi\cdots)$ denotes the quadratic and higher-order terms of the perturbations collectively.
We have chosen the gauge condition \cite{PhysRevD.85.064041}
\begin{equation}
\theta^{\mu\nu}_{\phantom{12},\mu}=0
\end{equation}
to simplify the field equation.
As a result of this condition, we have the conservation law \cite{PhysRevD.85.064041}
\begin{equation}
\tau^{\mu\nu}_{\phantom{ab},\mu}=0
\end{equation}

Substituting $M(\phi)=\frac12M''(\phi_0)\varphi^2$ into the scalar field equation \eqref{seq} and expanding this equation in the weak-field limit, we have 
\begin{equation}\label{weakscalar}
(\square_\eta-m_s^2) \varphi=-16\pi S,
\end{equation}
where the mass of the scalar field $m_s$ is given by
\begin{equation}\label{ms2}
m_s^2\equiv-\frac{\phi_0}{2\omega_0+3}M''(\phi_0),
\end{equation}
and the source $S$ is given by
\begin{align}
\begin{split}\label{s}
S=&-\frac{1}{16\pi}\left(\theta^{\mu\nu}\varphi_{,\mu\nu}+(\frac{1}{\phi_0}-\frac{\omega_1}{2\omega_0+3})\varphi_{,\alpha}\varphi^{,\alpha}-\frac12m_s^2\varphi\theta-(\frac{1}{\phi_0}+\frac{\omega_1}{2\omega_0+3})m_s^2\varphi^2\right)\\
&-\frac{1}{4\omega_0+6}\left(1-\frac{2\omega_1 \varphi}{2\omega_0+3}-\frac12\theta-\frac{\varphi}{\phi_0}\right)\left(T-2\phi\frac{\partial T}{\partial \phi}\right)\\
&+O(\theta^3,\theta^2\varphi,\theta\varphi^2\cdots).
\end{split}
\end{align}
The first line represents the field contribution to the source and the second line represents the  contribution from the stress-energy tensor of the particles.

\section{Motion of point particles}\label{eom}

Since we are going to calculate the gravitational waveform to the quadrupole order, we only need to solve the equation of motion of the point particles which generate the gravitational waves to Newtonian order. In this section, we rederive the equation of motion  obtained in \cite{PhysRevD.85.064041} to make this paper self-contained.
To Newtonian order, the tensor field equation \eqref{weak-t} becomes \cite{PhysRevD.85.064041}
\begin{equation}
\nabla^2\theta_{\mu\nu}=-\frac{16\pi}{\phi_0}T_{\mu\nu} 
\end{equation}
with
\begin{align}
\begin{split}
T_{00}=&\rho^*+O(\rho^*\epsilon^2)\\
T_{0i}=&O(\rho^*\epsilon)\\
T_{ij}=&O(\rho^* \epsilon^2)
\end{split}
\end{align}
where we have defined the density $\rho^*=\sum\limits_A m_A \delta^{(3)}({\bf x-x}_A)$ and $\epsilon$ is the typical velocity of the point particles.
Then, the solution for $\theta_{\mu\nu}$ to order $O(\epsilon^2)$ is \cite{PhysRevD.85.064041}
\begin{align}
\begin{split}\label{theta}
\theta_{00}&=\frac{4}{\phi_0}U,\\
\theta_{0i}&=0,\\
\theta_{ij}&=0,
\end{split}
\end{align}
where 
\begin{equation}
U\equiv \sum_A \frac{m_A}{|{\bf x-x}_A|}.
\end{equation}
To Newtonian order, the scalar field equation \eqref{weakscalar} becomes \cite{PhysRevD.85.064041}
\begin{equation}
(\nabla^2-m_s^2)\varphi=-\frac{8\pi}{2\omega_0+3}\sum_A m_A(1-2s_A)\delta^{(3)}({\bf x- x}_A)
\end{equation}
The solution is
\begin{equation}\label{phi}
\varphi = \frac{2}{2\omega_0+3}U_s
\end{equation}
with 
\begin{equation}
U_s\equiv\sum_A m_A(1-2s_A)\frac{e^{-m_s|{\bf x-x}_A|}}{|{\bf x-x}_A|}.
\end{equation}

Using the definition of $\theta_{\mu\nu}$ \eqref{thetamunu}, we obtain the metric perturbation to Newtonian order \cite{PhysRevD.85.064041}
\begin{align}
\begin{split}\label{hmunu}
h_{00}&=\frac{2}{\phi_0}U+\frac{2}{\phi_0(2\omega_0+3)}U_s,\\
h_{ij}&=\delta_{ij}\left[\frac{2}{\phi_0}U-\frac{2}{\phi_0(2\omega_0+3)}U_s\right],\\
h_{0i}&=0.
\end{split}
\end{align}

Using the Bianchi identity and the field equations \eqref{teq} and \eqref{seq}, we obtain the equation of motion
\begin{equation}
\nabla^\mu T_{\mu\nu}-\frac{\partial T}{\partial \phi}\partial_\nu \phi =0.
\end{equation}
We can also obtain this equation by using the invariant property of the matter action $S_m\left[g_{\mu\nu},\Psi_m\right]$ under diffeomorphisms \cite{PhysRevD.87.104020,wald1984general}.
Substituting the stress-energy tensor \eqref{set} for a single particle A into the above equation, we obtain the modified geodesic equation \cite{1975ApJ...196L..59E}
\begin{equation}
m_A(\phi) u^\mu_A\nabla_\mu u^\nu_A +\frac{d m_A(\phi)}{d \phi}(g^{\mu\nu}+u_A^\mu u_A^\nu)\nabla_\mu \phi=0.
\end{equation}
In the Newtonian limit, the modified geodesic equation becomes
\begin{equation}
\frac{d^2 x_A^i}{dt^2}+\Gamma^i_{00}+\frac{1}{\phi}\frac{d \ln m_A(\phi)}{d \ln \phi} \partial_i \phi=0
\end{equation}
where $\Gamma^i_{00}$ is the Christoffel symbol. It can be seen that the word line of a free particle with nonzero sensitivity is not a geodesic.
Substituting \eqref{phi} and \eqref{hmunu}  into the above equation yields
\begin{equation}\label{ddx}
\frac{d^2 {\bf x}_A}{dt^2}=-\frac{1}{\phi_0}\sum_B\frac{m_B{\bf r}_{AB}  }{r_{AB}^3}\left[1+(1-2s_A)(1-2s_B)(1+m_sr_{AB})\frac{e^{-m_s r_{AB}}}{2\omega_0+3}\right]
\end{equation}
with ${\bf r}_{AB}={\bf x}_A-{\bf x}_B$. This is the equation of motion of particle A to Newtonian order, which is consistent with (52) in \cite{PhysRevD.85.064041}.

We can use the above results to obtain the post-Newtonian expansion of the source $S$ \eqref{s}. Substituting \eqref{maphi} \eqref{phi} and \eqref{hmunu} into \eqref{set}, we have
\begin{equation}
-T+2\phi\frac{\partial T}{\partial \phi}= \rho^*\left[(1-2s)-3G(1-\xi)(1-2s)U-\frac12(1-2s)v^2+3(1-2s-\frac43 a_s)G\xi U_s+O(\epsilon)\right]
\end{equation}
where we have used the following parameters from \cite{PhysRevD.85.064041}
\begin{equation}\label{paras}
G\equiv\frac{1}{\phi_0}\frac{4+2\omega_0}{3+2\omega_0},\qquad \xi\equiv\frac{1}{2\omega_0+4},\qquad G(1-\xi)=\frac{1}{\phi_0}, \qquad a_s\equiv s^2+s'-\frac12 s.
\end{equation}
The body labels in $s$ and $a_s$ are omitted, since the delta function in $\rho^*$ will pick up the labels. We do not set the gravitational constant $G$ equal to 1, since it depends on the background scalar field $\phi_0$, which will evolve with the expansion of the  universe \cite{PhysRevD.96.064037}.
Therefore the post-Newtonian expansion of the  source $S$ is
\begin{equation}\label{exS}
S=S_C+S_F
\end{equation} 
where 
\begin{equation}\label{sc}
S_C=\frac{\rho^*}{4\omega_0+6}\left[(1-2s)-G(1-\xi)(1-2s)U-\frac12 (1-2s) v^2+ G\xi U_s \{3(1-2s-\frac43 a_s)-2(1-2s)(2+\lambda_1)\}+O(\epsilon^3)\right],
\end{equation}
\begin{equation}
S_F=-\frac{1}{16\pi}\left[G(1-\xi)(1-\lambda_1)\frac{4}{(2\omega_0+3)^2}\nabla\cdot(U_s\nabla U_s)-4m_s^2 G(1-\xi)(\lambda_1+2)\frac{1}{(2\omega_0+3)^2}U_s^2+4m_s^2G\xi U_s U+ \text{higher order}\right].
\end{equation}
The parameter $\lambda_1$ is given by
\begin{equation}
\lambda_1\equiv\frac{\omega_1\phi_0}{2\omega_0+3}.
\end{equation}
$S_C$ denotes the compact terms and $S_F$ originates from the second line in \eqref{s}, which represents the nonlinear field contribution.
In the limit $m_s=0$, the expansion of $S$ \eqref{exS} is consistent with equations (3.10a) and (3.10b) in \cite{Lang-BD}.
In the limit $\lambda_1=0$, the compact part $S_C$ \eqref{sc} agrees with the equation below equation (35) in \cite{PhysRevD.85.064041}. It is shown in the following section that only the first term in $S_C$ will contribute to the waveform at quadrupole order. Since $S_F$  is of higher PN order relative to $\rho^*$, we will ignore its contribution.

\section{Gravitational waves generated by the compact binary}\label{gw_waveform} 
\subsection{Time-domain GW waveforms}
In this section we will calculate the gravitational waveforms emitted by a compact binary system and its gravitational radiation power. 
Using the method of Green's function and multipole expansion, we obtain the quadrupole formula of the tensor wave \cite{PhysRevD.85.064041}
\begin{equation}
\theta^{ij}=\frac{2G(1-\xi)}{R}\frac{d^2}{dt^2}\sum_A m_A x_A^i x_A^j.
\end{equation}
Specializing to a two body system in the center of mass frame, we have \cite{PhysRevD.85.064041}
\begin{equation}\label{thetaij}
\theta^{ij}=\frac{4G(1-\xi)\mu}{R}(v^i v^j-\tilde{g}m\frac{r^i r^j}{r^3})
\end{equation}
where $m\equiv m_1+m_2$ is the total mass of the binary system. $\mu \equiv \frac{m_1 m_2}{m}$ is the reduced mass. $r^i\equiv x_1^i-x_2^i$ and $v^i\equiv v_1^i-v_2^i$ are the relative variables. $R$ is the coordinate distance of the field point relative to the center of mass. $\tilde{g}\equiv G(1-\xi)[1+\frac{1}{2\omega_0+3}(1-2s_1)(1-2s_2)(1+m_s r)e^{-m_s r}]$. We have used \eqref{ddx} to eliminate $\ddot{x}_A^i$.

To obtain the solution of the scalar wave equation \eqref{weakscalar}, we use the retarded Green's function $\mathcal{G}(x)$ which satisfy
\begin{equation}
(\square_\eta -m_s^2)\mathcal{G}(x)=-4\pi \delta^{(4)}(x)
\end{equation}
where $\delta^{(4)}(x)$ is the four dimensional delta function. The retarded Green's function is \cite{Poisson_2011}
\begin{equation}
\mathcal{G}(t,\mathbf{R})=\frac{\delta(t-R)}{R}-\Theta(t-R)\frac{m_s J_1(m_s\sqrt{t^2-R^2})}{\sqrt{t^2-R^2}},
\end{equation}
where $\Theta$ is the Heaviside function and $J_1$ is the Bessel function of the first kind and of order one. (For a detailed derivation, see section 12 in \cite{Poisson_2011}.)
The first term is supported on the future light cone of the source. The second term is supported within the future light cone. In the limit $m_s=0$, $\mathcal{G}(t,\mathbf{R})$ reduces to the Green's function of the wave operator $\square_\eta$.
Now the solution to the scalar wave equation \eqref{weakscalar} is \cite{PhysRevD.85.064041}
\begin{equation}
\varphi=\varphi_B+\varphi_m,
\end{equation}
where 
\begin{equation}
\varphi_B(t,\mathbf{R})=4 \int d^3 \mathbf{r}' dt'~\frac{S_C(t',\mathbf{r}')\delta(t-t'-|\mathbf{R-r'}|)}{|\mathbf{R-r'}|},
\end{equation}
\begin{equation}
\varphi_m(t,\mathbf{R})=-4\int d^3 \mathbf{r}' dt'~\Theta(t-t'-|\mathbf{R-r'}|)\frac{m_s S_C(t',\mathbf{r}')J_1(m_s\sqrt{(t-t')^2-|\mathbf{R-r'}|})}{\sqrt{(t-t')^2-|\mathbf{R-r'}|}}.
\end{equation}
The spatial integration is taken over the near zone. Note that we have discarded the contribution from $S_F$.
Taking the field point to be faraway, $R\gg |\mathbf{r}'|$, and keeping only the leading order $O(\frac{1}{R})$ part, we obtain the multipole expansion of the scalar wave \cite{PhysRevD.85.064041}
\begin{equation}
\varphi_B=\frac{4}{R}\sum_{k=0}^\infty \frac{1}{k!}\frac{\partial^k}{\partial t^k}\int d^3 \mathbf{r}' S_C(t-R,\mathbf{r}')(\mathbf{n}\cdot\mathbf{r}')^k,
\end{equation} 
\begin{equation}
\varphi_m=-\frac{4}{R}\sum_{k=0}^\infty \frac{1}{k!}\frac{\partial^k}{\partial t^k}\int d^3 \mathbf{r}'(\mathbf{n}\cdot\mathbf{r}')^k \int_0^\infty dz\frac{S_C(t-\sqrt{R^2+(\frac{z}{m_s})^2},\mathbf{r}')J_1(z)}{(1+(\frac{z}{m_s R})^2)^{(k+1)/2}},
\end{equation}
where $\mathbf{n}=\mathbf{R}/R$. Substituting the post-Newtonian expansion of the source $S_C$ \eqref{sc} into the above equations, we can obtain the post-Newtonian expansion of the scalar field. Since we only consider the quadrupole contribution to the tensor wave, we will keep only the terms up to order $O(\frac{mv^2}{R})$ in the scalar wave. Specializing to a two body system, we have
\begin{equation}\label{phib}
\varphi_B=\frac{2\alpha\mu}{R}\left[-2G(1-\xi)\frac{\Gamma'm}{r}-\frac12 \Gamma v^2-G\xi \beta \frac{m}{r}e^{-m_s r}-2\mathcal{S}(\mathbf{n}\cdot \mathbf{v})+\Gamma(\mathbf{n}\cdot \mathbf{v})^2-\Gamma\frac{\tilde{g}m}{r^3}(\mathbf{n}\cdot\mathbf{r})^2\right],
\end{equation}
\begin{equation}\label{phim}
\varphi_m=-\frac{2\alpha\mu}{R}\left[-2G(1-\xi)\Gamma'I_1[\frac{m}{r}]-\frac12\Gamma I_1[v^2]-G\xi\beta I_1[\frac{m}{r}e^{-m_s r}]-2\mathcal{S}I_2[\mathbf{n}\cdot \mathbf{v}]+\Gamma I_3[(\mathbf{n}\cdot \mathbf{v})^2]-\Gamma I_3[\frac{\tilde{g}m}{r^3}(\mathbf{n}\cdot\mathbf{r})^2]\right],
\end{equation}
where we have used the definition from \cite{PhysRevD.85.064041}
\begin{equation}
I_n[f(t)]\equiv \int_0^\infty dz ~\frac{f(t-Ru)J_1(z)}{u^n}
\end{equation}
with  $u\equiv\sqrt{1+(\frac{z}{m_s R})^2}$.
We have defined the parameters 
\begin{align}
\begin{split}
\alpha&\equiv\frac{1}{2\omega_0+3},\\
\Gamma'&\equiv 1-s_1-s_2,\\
\mathcal{S}&\equiv s_1-s_2,\\
\Gamma &\equiv \frac{(1-2s_1)m_2+(1-2s_2)m_1}{m},\\
\beta &\equiv (1-2s_2)[4 a_{s1}+(1-2s_1)(1+2\lambda_1)]+(1-2s_1)[4 a_{s2}+(1-2s_2)(1+2\lambda_1)].
\end{split}
\end{align}
The scalar dipole terms in \eqref{phib} and \eqref{phim} are proportional to $\mathcal{S}$. The scalar quadrupole terms are proportional to $\Gamma$. 
In the previous work \cite{PhysRevD.85.064041}, the coupling function $\omega(\phi)$ is set to be a constant. Comparing with equations (86) and (87) in \cite{PhysRevD.85.064041}, we find that the derivative of the coupling function $\omega(\phi)$ only modifies the monopole terms in the scalar wave.

The observational consequence of the gravitational waves in the long wavelength limit can be described by the geodesic deviation equation \cite{PhysRevLett.30.884,PhysRevD.8.3308}. The gravitational waves can influence the distance between the freely moving test particles. Assuming that the distance $\xi^i$ is small compared with the wavelength of the GWs and the test particles move slowly, the geodesic deviation equation becomes the approximate form  ${d^2}\xi^i/{dt^2}=-R_{0i0j}\xi^{j}$, where $R_{0i0j}$ is the Riemann tensor generated by the GWs. The GW field $\mathbf{h}_{ij}$ is defined by the Riemann tensor, $\partial^2 \mathbf{h}_{ij}/{\partial t^2}=-2 R_{0i0j}$. In a metric theory of gravity, there can be up to six polarizations of gravity \cite{PhysRevLett.30.884,PhysRevD.8.3308}. For a wave traveling in the $z$-direction, these polarizations become
\begin{equation}
 \mathbf{h}_{ij}(t)=\left(
    \begin{matrix}
    {h}_b+{h}_+ & {h}_{\times} & {h}_x \\
    {h}_{\times} & {h}_b-{h}_+ & {h}_y \\
    {h}_x & {h}_{y} & {h}_L
    \end{matrix}
    \right).
\end{equation}
Maggiore and Nicolis \cite{PhysRevD.62.024004} showed that the massive scalar field can induce two polarizations, the breathing polarization $h_b$ and the longitudinal polarization $h_L$. Now we will calculate the polarizations of the gravitational waves generated by a binary system in detail. For simplicity, we specialize to a quasicircular orbit. In this situation, the monopole terms in \eqref{phib} and \eqref{phim}
have no wavelike behavior. Therefore, we can discard these terms.

To linear order in the metric perturbation $h_{\mu\nu}$, the Riemann tensor is given by
\begin{equation}
R_{\mu\nu\alpha\beta}=\frac12(-h_{\mu\alpha,\nu\beta}+h_{\nu\alpha,\mu\beta}+h_{\mu\beta,\nu\alpha}-h_{\nu\beta,\mu\alpha}).
\end{equation}
Substituting the tensor wave \eqref{thetaij} and the scalar wave \eqref{phib} \eqref{phim} into the above equation, we have
\begin{equation}\label{r0i0j}
R_{0i0j}=-\frac12 \frac{\partial^2}{\partial t^2}[\theta_{ij}^{TT}-\frac{\varphi}{\phi_0}(\delta_{ij}-n_i n_j)-n_i n_j \frac{2\alpha \mu}{\phi_0 R}\int_0^\infty dz J_1(z)(\frac{1}{u^2}-1)\psi],
\end{equation}
where $\theta_{ij}^{TT}$ is the transverse-traceless part of $\theta_{ij}$ and
\begin{equation}
\psi\equiv \left[-\frac{2\mathcal{S}(\mathbf{n}\cdot \mathbf{v})}{u^2}+\frac{\Gamma(\mathbf{n}\cdot \mathbf{v})^2}{u^3}-\Gamma\frac{\tilde{g}m}{r^3}\frac{(\mathbf{n}\cdot\mathbf{r})^2}{u^3}\right]_{t-Ru}.
\end{equation}
We can read the four polarizations in massive Brans-Dicke theory from the Riemann tensor \eqref{r0i0j}.
\begin{equation}\label{hp}
h_+ =-4\delta\frac{(GM_c)^{5/3}}{R}~\omega^{2/3}~\frac{1+\cos^2\iota}{2}\cos(2\Phi)
\end{equation}
\begin{equation}\label{hc}
h_\times =-4\delta\frac{(GM_c)^{5/3}}{R}~\omega^{2/3}~\cos\iota ~\sin(2\Phi)
\end{equation}
\begin{equation}\label{hb}
h_b=-\frac{\varphi}{\phi_0}
\end{equation}
\begin{equation}\label{hl}
h_L=-\frac{2\alpha \mu}{\phi_0 R}\int_0^\infty dz J_1(z)(\frac{1}{u^2}-1)\psi
\end{equation}
where $\delta=(1-\xi)^{\frac53}[1+\alpha(1-2s_1)(1-2s_2)(1+m_s r)e^{-m_s r}]^{\frac23}$, $M_c = \mu^{3/5}m^{2/5}$ is the chirp mass, $\omega$ is the orbital frequency, $\Phi$ is the orbital phase and $\iota$ is the inclination angle between the binary orbital angular momentum and the line of sight. The scalar field $\varphi$ is given by
\begin{equation}
\varphi=\frac{2\alpha\mu}{R}[\Gamma v^2 \sin^2\iota \cos(2\Phi)-2\mathcal{S}v\sin\iota\cos(\Phi)]_{t-R}-\frac{2\alpha\mu}{R}\int_0^\infty dz J_1(z)\left[\frac{\Gamma v^2}{u^3}\sin^2\iota\cos(2\Phi)-\frac{2\mathcal{S}}{u^2}v \sin\iota\cos(\Phi)\right]_{t-Ru}
\end{equation}
We  have used the relations $\mathbf{n}\cdot\mathbf{v}=v\sin\iota\cos(\Phi)$ and $\mathbf{n}\cdot\mathbf{r}=r\sin\iota\sin(\Phi)$. We recall that $u=\sqrt{1+(\frac{z}{m_s R})^2}$. In Appendix \ref{int_hbhl}, we calculate the asymptotic behavior of the integrals in $h_b$ and $h_L$ when $R\to \infty$. After performing these integrals, the two scalar polarizations take the form
\begin{equation}\label{hb_waveform}
h_b=h_{b1}+h_{b2},
\end{equation} 
\begin{equation}\label{hb1}
h_{b1}=\frac{2G\xi \mu}{R}~ 2\mathcal{S}(\tilde{g}m)^{1/3}\sin\iota\ \omega^{1/3}\sqrt{1-\frac{m_s^2}{\omega^2}}\cos\left(\frac{m_s^2 R}{\sqrt{\omega^2-m_s^2}}+\Phi\right)\Theta(\omega-m_s)\Big|_{t-Ru_1},
\end{equation}
\begin{equation}\label{hb2}
h_{b2}=-\frac{2G\xi \mu}{R}~\Gamma(\tilde{g}m)^{2/3}\sin^2\iota\ \omega^{2/3}(1-\frac{m_s^2}{4\omega^2})\cos\left(\frac{m_s^2 R}{\sqrt{4\omega^2-m_s^2}}+2\Phi\right)\Theta(2\omega-m_s)\Big|_{t-Ru_2},
\end{equation}

and
\begin{equation}
h_L=h_{L1}+h_{L2},
\end{equation}
\begin{equation}
h_{L1}=\frac{m_s^2}{\omega^2}\frac{2G\xi \mu}{R} 2\mathcal{S}(\tilde{g}m)^{1/3}\sin\iota\ \omega^{1/3}\sqrt{1-\frac{m_s^2}{\omega^2}}\cos\left(\frac{m_s^2 R}{\sqrt{\omega^2-m_s^2}}+\Phi\right)\Theta(\omega-m_s)\Big|_{t-Ru_1},
\end{equation}
\begin{equation}\label{hl2}
h_{L2}=-\frac{m_s^2}{4\omega^2}\frac{2G\xi \mu}{R}\Gamma(\tilde{g}m)^{2/3}\sin^2\iota\ \omega^{2/3}(1-\frac{m_s^2}{4\omega^2})\cos\left(\frac{m_s^2 R}{\sqrt{4\omega^2-m_s^2}}+2\Phi\right)\Theta(2\omega-m_s)\Big|_{t-Ru_2},
\end{equation}
where $u_n=\left.{n\omega}/{\sqrt{n^2\omega^2-m_s^2}}\right|_{t-R}$. We have used the relation $v=(\tilde{g}m\omega)^{1/3}$ to eliminate $v$ and discarded the terms of order $O(\frac{e^{-R}}{R})$. Due to the existence of Heaviside function $\Theta$, a binary  system can radiate scalar waves only if the orbital frequency $\omega$ is high enough. The phase of the scalar wave satisfies the dispersion relation
\begin{equation}
\partial_\mu\left(\frac{m_s^2 R}{\sqrt{\omega^2-m_s^2}}+\Phi\right)\partial^\mu\left(\frac{m_s^2 R}{\sqrt{\omega^2-m_s^2}}+\Phi\right)=-m_s^2.
\end{equation}
If we ignore the time evolution of the orbital frequency $\omega$, then the phase takes a familiar form,
\begin{equation}
\frac{m_s^2 R}{\sqrt{\omega^2-m_s^2}}+\Phi=\omega t-kR+ \text{constant},
\end{equation}
where $k=\sqrt{\omega^2-m_s^2}$ is the wave number.
It can be seen that there is a simple linear relation between the breathing polarization $h_b$ and the longitudinal polarization $h_L$,
\begin{equation}
h_{L1}=\frac{m_s^2}{\omega^2} h_{b1}, \qquad h_{L2}=\frac{m_s^2}{4\omega^2} h_{b2}.
\end{equation}
This is a result of the linearized scalar wave equation \eqref{weakscalar} \cite{PhysRevD.62.024004,PhysRevD.98.083023}. Each of these two polarizations has two frequency modes. The lower frequency mode, proportional to $\mathcal{S}$, originates from the scalar dipole. The higher frequency mode, proportional to $\Gamma$, originates from the scalar quadrupole. 
In the limit $m_s=0$, $h_L=0$ and the waveforms of the others three polarizations become that of Brans-Dicke theory (section IIC in \cite{PhysRevD.95.124008}). 

We can use the above results to estimate the ratios between the amplitudes of different polarizations.
If the scalar field is heavy enough, $m_s\sim 10^{-16}\text{eV}$, then the parameter $\xi$ can be of order $10^{-2}$ \cite{PhysRevD.85.064041}. For a black hole-neutron star binary system with $m_{\text{BH}}=5 \text{M}_\odot$ and $m_{\text{NS}}=1.4\text{M}_\odot$, when the frequency of the tensor wave is about 100Hz, the ratio of the amplitude of the breathing polarization to that of the plus polarization is about
\begin{equation}
\frac{|h_b|}{|h_+|}\sim 10^{-2},
\end{equation}
and the ratio of the amplitude of the longitudinal polarization to that of the breathing polarization is about
\begin{equation}
\frac{|h_L|}{|h_b|}\sim 10^{-7}.
\end{equation}

The signal received by a GW detector is given by the response function \cite{Poisson2014}
\begin{equation}
h(t)=F_+h_++F_\times h_\times+F_b h_b+F_L h_L
\end{equation}
where the detector antenna pattern functions $F_A ~(A=+,\times,b,L)$ depend on the geometry and orientation of the detector. For the explicit expressions of the antenna pattern functions, please refer to \cite{PhysRevD.92.063013} and section 13.4.3 in \cite{Poisson2014}.
The expressions of the tensor waveforms \eqref{hc} and \eqref{hp} have been obtained in \cite{PhysRevD.85.122005}. The expressions of the scalar waveforms \eqref{hb_waveform}-\eqref{hl2} are the new results. These results are helpful in the future search for the massive scalar field.

\subsection{GW radiation power}
Following \cite{PhysRevD.85.064041}, we derive the GW radiation power of a binary system in massive Brans-Dicke theory. The GW energy-momentum tensor is \cite{Saffer_2018}
\begin{equation}
t_{\alpha\beta}=\langle\frac{\phi_0}{32\pi}\theta_{TT,\alpha}^{\mu\nu}\theta_{\mu\nu,\beta}^{TT}+\frac{\phi_0}{16\pi}(3+2\omega_0)\frac{\varphi_{,\alpha}\varphi_{,\beta}}{\phi_0^2}\rangle
\end{equation}
where  $\langle\cdots\rangle$ represents the average over several wavelengths of GWs.
The GW radiation power in massive Brans-Dicke theory is \cite{maggiore2008gravitational}
\begin{equation}
\frac{dE_{GW}}{dt}=\int d \Omega~ R^2~ t^{0R}=\int d \Omega~ R^2\langle\frac{\phi_0}{32\pi}\dot{\theta}^{TT}_{ij}\dot{\theta}^{TT}_{ij}-\frac{\phi_0}{16\pi}(3+2\omega_0)\frac{\varphi_{,0}\varphi_{,R}}{\phi_0^2}\rangle,
\end{equation}
where $d\Omega$ denotes the solid angle element. If both the scalar and the tensor gravitational waves are massless, then $t^{0R}=t^{00}+O(\frac{1}{R^3})$ \cite{maggiore2008gravitational}. However, when the scalar field is massive,  $t^{0R}\neq t^{00}+O(\frac{1}{R^3})$. Will only studies massless scalar waves in \cite{will1993theory,*will_2018}. The radiation power due to the tensor field is given by \cite{PhysRevD.85.064041}
\begin{equation}
\frac{dE_t}{dt}=\int d\Omega ~R^2\langle\frac{\phi_0}{32\pi}\dot{\theta}^{TT}_{ij}\dot{\theta}^{TT}_{ij}\rangle=\frac{\phi_0}{16\pi}\int d\Omega~ R^2\langle\dot{h}^2_++\dot{h}^2_\times\rangle=\frac{32}{5}\phi_0\delta^2(GM_c\omega)^{10/3},
\end{equation}
where we have used the  expressions \eqref{hp} and \eqref{hc} of the two tensor polarizations. The radiation power due to the scalar field is given by
\begin{equation}\label{des}
\frac{dE_s}{d t}=-\frac{3+2\omega_0}{16\pi \phi_0} R^2 \int d \Omega \langle\varphi_{,0}\varphi_{,R}\rangle.
\end{equation}
Using \eqref{phib} and \eqref{phim}, we have
\begin{equation}\label{phi0}
\varphi_{,0}=\frac{2\alpha\tilde{g}m\mu}{R}\left[-4\Gamma\left(\frac{(\mathbf{n}\cdot\mathbf{r})(\mathbf{n}\cdot\mathbf{v})}{r^3}-I_3[\frac{(\mathbf{n}\cdot\mathbf{r})(\mathbf{n}\cdot\mathbf{v})}{r^3}]\right)+2\mathcal{S}\left(\frac{\mathbf{n}\cdot\mathbf{r}}{r^3}-I_2[\frac{\mathbf{n}\cdot\mathbf{r}}{r^3}]\right)\right],
\end{equation}
\begin{equation}\label{phiR}
\varphi_{,R}=-\frac{2\alpha\tilde{g}m\mu}{R}\left[-4\Gamma\left(\frac{(\mathbf{n}\cdot\mathbf{r})(\mathbf{n}\cdot\mathbf{v})}{r^3}-I_4[\frac{(\mathbf{n}\cdot\mathbf{r})(\mathbf{n}\cdot\mathbf{v})}{r^3}]\right)+2\mathcal{S}\left(\frac{\mathbf{n}\cdot\mathbf{r}}{r^3}-I_3[\frac{\mathbf{n}\cdot\mathbf{r}}{r^3}]\right)\right].
\end{equation}
Note that $\varphi_{,0}\neq-\varphi_{,R}+O(\frac{1}{R^2})$ for the massive scalar wave. 
In section VIIB in \cite{PhysRevD.85.064041}, the energy loss rate due to the scalar field is given by $\frac{dE}{d t}=-\frac{3+2\omega_0}{16\pi \phi_0} R^2 \int d \Omega \langle\varphi_{,0}\varphi_{,0}\rangle$, which assumes  $\varphi_{,0}=-\varphi_{,R}+O(\frac{1}{R^2})$.  Actually, the GW waveforms in \cite{PhysRevD.85.122005} are based on this radiation power.  

Substituting \eqref{phi0} and \eqref{phiR} into \eqref{des} yields
\begin{align}
\begin{split}
\int d\Omega\langle\varphi_{,0}\varphi_{,R}\rangle=&-\left(\frac{2\alpha\tilde{g}m\mu}{R}\right)^2\int d\Omega\langle 4\mathcal{S}^2\left(\frac{\mathbf{n}\cdot\mathbf{r}}{r^3}-I_2[\frac{\mathbf{n}\cdot\mathbf{r}}{r^3}]\right)\left(\frac{\mathbf{n}\cdot\mathbf{r}}{r^3}-I_3[\frac{\mathbf{n}\cdot\mathbf{r}}{r^3}]\right)\\
&+16 \Gamma^2\left(\frac{(\mathbf{n}\cdot\mathbf{r})(\mathbf{n}\cdot\mathbf{v})}{r^3}-I_3[\frac{(\mathbf{n}\cdot\mathbf{r})(\mathbf{n}\cdot\mathbf{v})}{r^3}]\right)\left(\frac{(\mathbf{n}\cdot\mathbf{r})(\mathbf{n}\cdot\mathbf{v})}{r^3}-I_4[\frac{(\mathbf{n}\cdot\mathbf{r})(\mathbf{n}\cdot\mathbf{v})}{r^3}]\right)\rangle,
\end{split}
\end{align}
\begin{align}
\begin{split}
\frac{dE_s}{dt}=\frac{G\xi\tilde{g}^2m^2\mu^2}{r^4}\{&\frac{16}{15}\Gamma^2v^2[1-\cos(2\omega R)(C_3(R;2\omega)+C_4(R;2\omega))-\sin(2\omega R)(S_3(R;2\omega)+S_4(R;2\omega))\\
&+C_3(R;2\omega)C_4(R;2\omega)+S_3(R;2\omega)S_4(R;2\omega)]\\
&+\frac{4}{3}\mathcal{S}^2[1-\cos(\omega R)(C_2(R;\omega)+C_3(R;\omega))-\sin(\omega R)(S_2(R;\omega)+S_3(R;\omega))\\
&+C_2(R;\omega)C_3(R;\omega)+S_2(R;\omega)S_3(R;\omega)]\},
\end{split}
\end{align}
where we have used the following integrals from \cite{PhysRevD.85.064041}
\begin{equation}
C_n(R;\omega)=\int_0^\infty dz \cos(\omega R u)\frac{J_1(z)}{u^n}, \qquad S_n(R;\omega)=\int_0^\infty dz \sin(\omega R u)\frac{J_1(z)}{u^n}.
\end{equation}
We recall that $u\equiv\sqrt{1+(\frac{z}{m_s R})^2}$. In the integrals $C_n$ and $S_n$, the orbital frequency $\omega$ is assumed to be a constant.
This assumption is also used when we calculated the tensor radiation power $dE_t/dt$.
The asymptotic expansion of $C_n$ and $S_n$ for $R\to \infty$ is \cite{PhysRevD.85.064041}
\begin{equation}
C_n(R;\omega)\sim\begin{cases}
\cos(\omega R)-\left(\frac{\sqrt{\omega^2-m_s^2}}{\omega}\right)^{n-1}\cos(R\sqrt{\omega^2-m_s^2}), & \omega>m_s\\
\cos(\omega R)-\left(\frac{\sqrt{\omega^2-m_s^2}}{\omega}\right)^{n-1}e^{-R\sqrt{m_s^2-\omega^2}}\cos\frac{(n-1)\pi}{2},&  \omega<m_s
\end{cases}
\end{equation}
\begin{equation}
S_n(R;\omega)\sim\begin{cases}
\sin(\omega R)-\left(\frac{\sqrt{\omega^2-m_s^2}}{\omega}\right)^{n-1}\sin(R\sqrt{\omega^2-m_s^2}), & \omega>m_s\\
\sin(\omega R)-\left(\frac{\sqrt{\omega^2-m_s^2}}{\omega}\right)^{n-1}e^{-R\sqrt{m_s^2-\omega^2}}\sin\frac{(n-1)\pi}{2},&  \omega<m_s
\end{cases}
\end{equation}
(For the details to obtain the asymptotic expansion of these two integrals\footnote{There is a typo in equation (B12) in \cite{PhysRevD.85.064041}, where $\frac{i^{n-1}-(-i)^{n-1}}{2}$ should be replaced by $\frac{i^{n-1}-(-i)^{n-1}}{2i}$.}, see Appendix B in \cite{PhysRevD.85.064041}.)  Using these results, we have the scalar radiation power
\begin{equation}\label{des1}
\frac{dE_s}{d t}=\frac{G\xi \tilde{g}^2 m^2 \mu^2}{r^4}\left[\frac{16}{15}\Gamma^2 v^2 \left(\frac{\sqrt{4\omega^2-m_s^2}}{2\omega}\right)^5\Theta(2\omega-m_s)+\frac{4}{3}\mathcal{S}^2\left(\frac{\sqrt{\omega^2-m_s^2}}{\omega}\right)^3\Theta(\omega-m_s)\right].
\end{equation}
The first term represents the scalar quadrupole radiation and the second term represents the scalar dipole radiation.
In the limit $m_s=0$, the radiation power is consistent with that of Brans-Dicke theory, equation (16) in \cite{PhysRevD.95.124008}. 
Actually, we can also use the waveform of the breathing polarization  to obtain the scalar radiation power.
Using \eqref{hb} and \eqref{des}, we have 
\begin{equation}
\frac{dE_s}{d t}=-\frac{(3+2\omega_0)\phi_0}{16\pi } R^2 \int d \Omega \langle h_{b,0} h_{b,R}\rangle.
\end{equation}
Substituting \eqref{hb_waveform}-\eqref{hb2} into the above equation yields \eqref{des1}. 

Since the  waveform $h_b$ begins at the dipole order, which is of -0.5PN order relative to the quadrupole term, the 0.5PN term(s) in $h_b$ will contribute to the scalar radiation power at the quadrupole order. However, the 0.5PN contribution to $h_b$ is beyond the scope of this paper.

\subsection{Frequency-domain GW waveforms}
In GW data analysis, one often works with the Fourier transforms of the GW waveforms. In order to obtain the frequency-domain GW waveforms, we need the time evolution of the orbital frequency $\omega$. 
Using the energy balance condition $\frac{dE}{dt}=-\frac{dE_{GW}}{dt}$ with $E=-\frac12\mu v^2=-\frac12\mu(\tilde{g}m\omega)^{\frac23}$, we have
the time derivative of the orbital frequency
\begin{align}
\begin{split}
\dot{\omega}=&\frac{96}{5}(GM_c(1-\xi))^{\frac53}[1+\alpha(1-2s_1)(1-2s_2)(1+m_s r)e^{-m_s r}]^{\frac23}\omega^{\frac{11}{3}}\\
&+3G\xi \mu\omega^3[\frac{16}{15}\Gamma^2(\tilde{g}m\omega)^{\frac23}(1-\frac{m_s^2}{4\omega^2})^{\frac52}\Theta(2\omega-m_s)+\frac43\mathcal{S}^2(1-\frac{m_s^2}{\omega^2})^{\frac32}\Theta(\omega-m_s)].
\end{split}
\end{align}

We will only consider the Fourier transforms of the tensor polarizations, since the dominant observational constraint comes from these polarizations \cite{Niu_2020}.
The Fourier transforms of the plus polarization $h_+$ is given by
\begin{equation}
\tilde{h}_+(f)=\int h_+(t)e^{i2\pi ft}dt.
\end{equation}
Using the stationary phase approximation, we have
\begin{equation}
\tilde{h}_+(f)=-2\delta \frac{(GM_c)^{\frac53}}{R}\frac{1+\cos^2\iota}{2}\omega(t_*)^{\frac23}\sqrt{\frac{\pi}{\dot{\omega}(t_*)}}e^{i\Psi_+},
\end{equation}
where $t_*$ is determined by
\begin{equation}
\omega(t_*)=\pi f,
\end{equation}
and
\begin{equation}\label{phase}
\Psi_+=2\pi fR-2\Phi(t_*)+2\pi ft_*-\frac{\pi}{4}.
\end{equation}
We use the relation
\begin{equation}\label{int_in_phase}
2\pi ft_*-2\Phi(t_*)=\int_\infty^{\pi f}\frac{2\pi f-2 \omega}{\dot{\omega}}d\omega+2\pi f t_c-2\Phi_c
\end{equation}
to express $t_*$ in terms of $f$, where $t_c$ is determined by $\omega(t_c)=\infty$ and $\Phi_c=\Phi(t_c)$.
We will evaluate the integral in different cases. When the scalar mass is light $m_s\ll \pi f$, the integral becomes
\begin{align}
\begin{split}
&\int_\infty^{\pi f}\frac{2\pi f-2 \omega}{\dot{\omega}}d\omega\\
=&\int_\infty^{\pi f}d\omega(2\pi f-2\omega)\frac{5}{96}(GM_c)^{-\frac53}\omega^{-\frac{11}{3}}\left\{1+\frac53\xi-\frac23\alpha(1-2s_1)(1-2s_2)\right.\\
&\left.-\frac{5}{32}\xi\left[\frac{16}{15}\Gamma^2(1-\frac{5}{8}\frac{m_s^2}{\omega^2})+\frac43\frac{\mathcal{S}^2}{(Gm\omega)^\frac23}(1-\frac32\frac{m_s^2}{\omega^2})\right]\right\}\\
=&\frac{3}{128}(GM_c\pi f)^{-\frac53}\left\{1+\frac53\xi-\frac23\alpha(1-2s_1)(1-2s_2)-\frac{20}{3}\xi\left[\frac{1}{11648}\frac{\mathcal{S}^2}{(Gm\pi f)^{\frac23}}\left(208-105\left(\frac{m_s}{\pi f}\right)^2\right)\right.\right.\\
&\left.\left.+\frac{\Gamma^2}{6160}\left(154-25\left(\frac{m_s}{\pi f}\right)^2\right)\right]\right\}.
\end{split}
\end{align}
For a light scalar mass ($m_s<2.5\times10^{-20}~\text{eV}$), the Cassini spacecraft has constrained $\omega_0$ to be larger than 40~000 \cite{PhysRevD.85.064041}, i.e. $\xi<10^{-5}$. Therefore, we retain only terms to the order $O(\frac{1}{\omega_0})$ in the above equation. Then, the phase $\Psi_+$ in the light scalar mass situation becomes
\begin{align}\label{PsiP3}
\begin{split}
\Psi_+=&2\pi f(R+t_c)-2\Phi_c-\frac{\pi}{4}+\frac{3}{128}(GM_c\pi f)^{-\frac53}\left\{1+\frac53\xi-\frac23\alpha(1-2s_1)(1-2s_2)
\right.\\
&\left.-\frac{20}{3}\xi\left[\frac{1}{11648}\frac{\mathcal{S}^2}{(Gm\pi f)^{\frac23}}\left(208-105\left(\frac{m_s}{\pi f}\right)^2\right)+\frac{\Gamma^2}{6160}\left(154-25\left(\frac{m_s}{\pi f}\right)^2\right)\right]\right\}.
\end{split}
\end{align}
The Fourier transforms of the tensor polarizations in the light scalar mass situation are
\begin{align}
\begin{split}
\tilde{h}_+(f)=&-\delta\frac{(GM_c)^{\frac56}}{R}\frac{1+\cos^2\iota}{2}\left(\frac{5\pi}{24}\right)^{\frac12}(\pi f)^{-\frac76}\left\{1+\frac56\xi-\frac13\alpha(1-2s_1)(1-2s_2)\right.\\
&\left.-\frac{5}{64}\xi\left[\frac{16}{15}\Gamma^2\left(1-\frac52\left(\frac{m_s}{2\pi f}\right)^2\right)+\frac43\frac{\mathcal{S}^2}{(Gm\pi f)^\frac23}\left(1-\frac32\left(\frac{m_s}{\pi f}\right)^2\right)\right] \right\}e^{i\Psi_+},
\end{split}
\end{align}
\begin{align}\label{hcf}
\begin{split}
\tilde{h}_\times(f)=&-\delta\frac{(GM_c)^{\frac56}}{R}\cos\iota\left(\frac{5\pi}{24}\right)^{\frac12}(\pi f)^{-\frac76}\left\{1+\frac56\xi-\frac13\alpha(1-2s_1)(1-2s_2)\right.\\
&\left.-\frac{5}{64}\xi\left[\frac{16}{15}\Gamma^2\left(1-\frac52\left(\frac{m_s}{2\pi f}\right)^2\right)+\frac43\frac{\mathcal{S}^2}{(Gm\pi f)^\frac23}\left(1-\frac32\left(\frac{m_s}{\pi f}\right)^2\right)\right] \right\}e^{i\Psi_\times},
\end{split}
\end{align}
with $\Psi_\times=\Psi_++\frac{\pi}{2}$ and $\Psi_+$ is given by \eqref{PsiP3}. In the limit $\omega_0\to \infty$, $\xi=\alpha=0$ and $\delta=1$, the expressions of $\tilde{h}_+(f)$ and $\tilde{h}_\times(f)$ reduce to that of GR.

When the scalar mass is  $m_s$ of order $\pi f$, the experiments do not exclude that $\omega_0$ is of order one \cite{PhysRevD.85.064041}. Therefore, we cannot linearize in $\xi$. The integral in the phase $\Psi_+$ becomes
\begin{align}
\begin{split}\label{hmphase}
&\int_\infty^{\pi f}\frac{2\pi f-2 \omega}{\dot{\omega}}d\omega\\
=&\int_\infty^{\pi f}d\omega~(2\pi f-2\omega)\frac{5}{96}(GM_c(1-\xi))^{-\frac53}\omega^{-\frac{11}{3}}\left\{1-\frac23\frac{\xi}{1-\xi}\Gamma^2(1-\frac{m_s^2}{4\omega^2})^\frac52\Theta(2\omega-m_s)\right.\\
&\left.-\frac{5}{24}(Gm\omega)^{-\frac23}\frac{\xi}{(1-\xi)^\frac53}\mathcal{S}^2(1-\frac{m_s^2}{\omega^2})^\frac32\Theta(\omega-m_s)\right\}\\
=&\frac{3}{128}[GM_c(1-\xi)\pi f]^{-\frac{5}{3}}-\frac{5}{144}\Theta(2\pi f-m_s)(GM_c)^{-\frac53}\frac{\xi}{(1-\xi)^\frac{3}{8}}\Gamma^2~\int_{\frac{m_s}{2}}^{\pi f}d\omega~(2\pi f-2\omega)\omega^{-\frac{11}{3}}(1-\frac{m_s^2}{4\omega^2})^{\frac52}\\
&-\frac{25}{2304}\Theta(\pi f-m_s)(GM_c)^{-\frac53}(Gm)^{-\frac23}\frac{\xi}{(1-\xi)^{\frac{10}{3}}}\mathcal{S}^2~\int_{m_s}^{\pi f}d\omega~(2\pi f-2\omega)\omega^{-\frac{13}{3}}(1-\frac{m_s^2}{\omega^2})^{\frac32}
\end{split}
\end{align}
We have discarded the constant terms which are independent of the frequency $f$, since they can be absorbed into $\Phi_c$. 
The two integrals in the last two lines can be expressed in terms of hypergeometric functions (see Appendix \ref{two integrals}). Using the results of these two integrals, we can obtain the frequency domain waveforms of $\tilde{h}_+(f)$ and $\tilde{h}_\times(f)$. However, in order to have a better understanding of this result, we calculate these integrals in two limits and use \eqref{phase} and \eqref{int_in_phase} to obtain the phase $\Psi_+$ of the waveforms in these two limiting cases.

In the limit $m_s\to \pi f$, the phase $\Psi_+$ becomes
\begin{align}
\begin{split}\label{PsiP1}
\Psi_+=&2\pi f(R+t_c)-2\Phi_c-\frac{\pi}{4}+\frac{3}{128}[GM_c(1-\xi)\pi f]^{-\frac53}\left\{1-\frac{40}{27}\frac{\xi}{1-\xi}\Gamma^2[0.103+0.662(1-\frac{m_s}{\pi f})]\right.\\
&\left.-\frac{40\sqrt{2}}{189}\Theta(\pi f-m_s)\frac{\xi}{(1-\xi)^{\frac53}}\mathcal{S}^2(Gm\pi f)^{-\frac23}(1-\frac{m_s}{\pi f})^{\frac72}\right\}.
\end{split}
\end{align}

In the limit $m_s\to 2\pi f$, the phase $\Psi_+$ becomes
\begin{equation}\label{PsiP2}
\Psi_+=2\pi f(R+t_c)-2\Phi_c-\frac{\pi}{4}+\frac{3}{128}[GM_c(1-\xi)\pi f]^{-\frac53}\left\{1-\frac{1280\sqrt{2}}{1701}\Theta(2\pi f-m_s)\frac{\xi}{1-\xi}\Gamma^2(1-\frac{m_s}{2\pi f})^{\frac92}\right\}.
\end{equation}
Therefore, in these two limits the frequency domain waveforms are
\begin{align}\label{hpfH}
\begin{split}
\tilde{h}_+(f)=&-\delta\frac{(GM_c)^{\frac56}}{R}\frac{1+\cos^2\iota}{2}\left(\frac{5\pi}{24}\right)^{\frac12}(\pi f)^{-\frac76}(1-\xi)^{-\frac56}\left\{1+\frac23\frac{\xi}{1-\xi}\Gamma^2\left(1-\frac{m_s^2}{4\pi^2 f^2}\right)^\frac52\Theta(2\pi f-m_s)\right.\\
&\left.+\frac{5}{24}(Gm\pi f)^{-\frac23}\frac{\xi}{(1-\xi)^\frac53}\mathcal{S}^2\left(1-\frac{m_s^2}{\pi^2 f^2}\right)^\frac32
\Theta(\pi f-m_s)  \right\}^{-\frac12}e^{i\Psi_+},
\end{split}
\end{align}

\begin{align}\label{hcfH}
\begin{split}
\tilde{h}_\times(f)=&-\delta\frac{(GM_c)^{\frac56}}{R}\cos\iota\left(\frac{5\pi}{24}\right)^{\frac12}(\pi f)^{-\frac76}(1-\xi)^{-\frac56}\left\{1+\frac23\frac{\xi}{1-\xi}\Gamma^2\left(1-\frac{m_s^2}{4\pi^2 f^2}\right)^\frac52\Theta(2\pi f-m_s)\right.\\
&\left.+\frac{5}{24}(Gm\pi f)^{-\frac23}\frac{\xi}{(1-\xi)^\frac53}\mathcal{S}^2\left(1-\frac{m_s^2}{\pi^2 f^2}\right)^\frac32
\Theta(\pi f-m_s)  \right\}^{-\frac12}e^{i\Psi_\times},
\end{split}
\end{align}
where $\Psi_\times=\Psi_++\frac{\pi}{2}$ and $\Psi_+$ is given by \eqref{PsiP1} or \eqref{PsiP2}. We recall that $\delta$ is defined below \eqref{hl}.
It can be seen that when  $m_s\to \infty$, $\delta=(1-\xi)^{5/3}$ and the expressions of $\tilde{h}_+(f)$ and $\tilde{h}_\times(f)$  reduce to that of GR, except for replacing the chirp mass  $M_c$ with $M_c(1-\xi)$.

The sensitivity of a black hole in massive Brans-Dicke theory is $s_{\text{BH}}=\frac12$ \cite{PhysRevD.98.083023}. As a result, for a binary black hole system, $\Gamma=\mathcal{S}=0$ and the waveforms are identical to that of GR apart from the replacement  $M_c\to M_c(1-\xi)$ which are the same as that in the limit $m_s\to \infty$. This is because in both cases the binary system has no scalar radiation.

\section{Parametrized post-Einsteinian parameters and observational constraints}\label{ppeob}
The parametrized post-Einsteinian (ppE) framework is a waveform model to describe the GWs emitted by a binary system on a quasicircular orbit in metric theories of  gravity. In the original ppE framework,  Yunes and Pretorius \cite{PhysRevD.80.122003} propose that the GW waveform of a binary during the inspiral is
$\tilde{h}(f)=\tilde{h}_{\rm GR} (f)\left(1+\alpha_{\text{ppe}}(G M_c\pi f)^\frac{a}{3}\right)e^{i \beta_{\text{ppe}}( G M_c \pi f)^\frac{b}{3}},$
where $\tilde{h}_{\rm GR} (f)$ is the GR Fourier waveform and $(\alpha_{\text{ppe}},\beta_{\text{ppe}},a,b)$ are the four ppE parameters that describe the non-GR correction to the GW amplitude and phase. Note that the original ppE framework only consider the two tensor polarizations $h_+$ and $h_\times$.
Clearly, this parametrization cannot describe the waveforms in the previous section. We need a more general framework
\begin{equation}
\tilde{h}(f)=\tilde{h}_{\rm GR} (f)\left(1+\sum_j\alpha_j(G M_c\pi f)^\frac{a_j}{3}\right)e^{i \sum_j \beta_j( G M_c \pi f)^\frac{b_j}{3}}.
\end{equation}
From \eqref{PsiP3}-\eqref{hcf}, we obtain four sets of the ppE parameters of massive Brans-Dicke theory in the light scalar mass situation

\begin{align}
\alpha_1&=-\frac{5}{48}\xi\mathcal{S}^2\eta^{2/5}, & \quad \beta_1&=-\frac{5}{1792}\xi\mathcal{S}^2\eta^{2/5}, \quad
& a_1&=-2, &\quad b_1&=-7~,\\
\alpha_2&=\frac{5}{32}\xi\mathcal{S}^2\eta^{2/5}(GM_cm_s)^2, & \quad \beta_2&=\frac{75}{53248}\xi\mathcal{S}^2\eta^{2/5}(GM_cm_s)^2, \quad
&a_2&=-8, &\quad b_2&=-13~,\\
\alpha_3&=-\frac{1}{12}\xi\Gamma^2, &\quad \beta_3&=-\frac{1}{256}\xi\Gamma^2, \quad
&a_3&=0, &\quad b_3&=-5~,\\
\alpha_4&=\frac{5}{96}\xi\Gamma^2(GM_cm_s)^2,& \quad \beta_4&=\frac{25}{39424}\xi\Gamma^2(GM_cm_s)^2, \quad
&a_4&=-6, &\quad b_4&=-11~,
\end{align}
The first two sets of ppE parameter correspond to the scalar dipole radiation and the last two sets correspond to the scalar quadrupole radiation. 
The first set of ppE parameter $(\alpha_1,\beta_1,a_1,b_1)$ is consistent with the ppE parameters of massless Brans-Dicke theory obtained in the  previous work \cite{Yunes-BD}.

Chamberlain and Yunes have studied the observational constraints on $\beta_{\text{ppe}}$ by the future ground-based GW detectors, the LIGO-class expansions A+, Voyager, Cosmic Explorer and the Einstein Telescope \cite{PhysRevD.96.084039}. They considered the GWs emitted by a black hole-neutron star system with $m_{\text{BH}}=5 \text{M}_\odot$ and $m_{\text{NS}}=1.4\text{M}_\odot$ at the distance 150 Mpc. Assuming that the detection is consistent with GR, they obtained the constraints on $\beta_{\text{ppe}}$ listed in the second column in Table \ref{bounds}. The typical value of the sensitivity of a neutron star is $s_{\rm NS}=0.2$ \cite{Damour_1992}. Using the expressions of ppE parameters of massive Brans-Dicke theory, we obtain the constraints on the parameters in this theory listed in the third column in Table \ref{bounds}.  

\begin{table}[h]
\caption{The projected constraints on $\beta_{\text{ppe}}$ and the parameters of massive Brans-Dicke theory as a function of  exponent parameter $b$ \cite{PhysRevD.96.084039}. $m_{20}=10^{-20}~\text{eV}$.}\label{bounds}
\begin{tabular}{ccc}
\hline\hline
$b$ & upper bound on $|\beta_{\text{ppe}}|$ & parameter constraint\\
\hline
-5 & $3.48\times 10^{-4}$ & $\xi<0.41$\\
\hline
-7 & $2.88\times 10^{-8}$ & $\xi<2.3\times 10^{-4}$\\
\hline
-11 & $1.88\times 10^{-13}$ & $\xi(m_s/m_{20})^2<4.9\times 10^{10}$\\
\hline
-13 & $6.95\times 10^{-16}$ & $\xi(m_s/m_{20})^2<4.0\times 10^8$\\
\hline\hline
\end{tabular}
\end{table}

However, when the scalar mass $m_s$ is comparable to $\pi f$, the Fourier waveforms \eqref{hpfH} and \eqref{hcfH} cannot be described by the ppE framework.
Therefore, there is no available constraint on the parameters of massive Brans-Dicke theory in this situation.

Let us now turn to apply our result to specific models.
\subsection{DEF model}
This scalar-tensor theory is proposed by Damour and Esposito-Far\`{e}se (DEF) to study the spontaneous scalarization of neutron stars \cite{PhysRevLett.70.2220,PhysRevLett.124.221104}. The action is given in the Einstein frame

\begin{equation}\label{action2}
S_E=\int d^4 x\sqrt{-g_*}\left[\frac{1}{16\pi G_*} R_*-\frac12~ \partial_\mu\phi_*\partial^\mu\phi_*-V(\phi_*)\right]+S_m\left[A^2(\phi_*)g^*_{\mu\nu},\Psi_m\right],
\end{equation}
where $R_*$ is the Ricci scalar of the Einstein frame metric $g^*_{\mu\nu}$. $g_*$ is the determinant of $g^*_{\mu\nu}$. $G_*$ denotes the gravitational constant. $V(\phi_*)$ is the scalar potential and $A(\phi_*)$ is the conformal coupling function. Using the conformal transformation $g_{\mu\nu}=A^2(\phi_*)g^*_{\mu\nu}$ between the Jordan frame and the Einstein frame, we  obtain the relation between these two frames
\begin{equation}\label{relation}
\phi=\frac{1}{G_* A^2(\phi_*)},\quad \frac{4\pi G_*}{2\omega(\phi)+3}= \left(\frac{d\ln A(\phi_*)}{d \phi_*}\right)^2, \quad M(\phi)=-\frac{16\pi V(\phi_*)}{A^4(\phi_*)}
\end{equation}
The coupling function in the DEF model is given by \cite{PhysRevLett.70.2220}
\begin{equation}
A(\phi_*)=\exp\left[\frac14\beta_*\left(\frac{\phi_*}{M_p}\right)^2\right] ,
\end{equation}
where $\beta_*$ is a constant and $M_p=1/\sqrt{8\pi G_*}$. The scalar field is massless in the original DEF model. Then, Ramazano\v{g}lu and  Pretorius \cite{PhysRevD.93.064005} extend this model to study the spontaneous scalarization of neutron stars with a massive scalar field. The potential term in the extended DEF model is given by
\begin{equation}
V(\phi_*)=\frac12m_*^2\phi_*^2,
\end{equation}
with $m_*$ the scalar mass in the Einstein frame. In \cite{PhysRevD.93.064005} the background scalar field is the minimum of the potential $V(\phi_*)$, $\phi_*=0$. 
From \eqref{relation}, we have 
\begin{equation}
\omega_0\to+\infty.
\end{equation}
Combining with \eqref{ms2} and \eqref{paras}, we obtain
\begin{equation}
m_s=0,\quad \xi=0.
\end{equation}
As a result of the special background scalar field value, the extended DEF model can satisfy the constraints in Table \ref{bounds}.

Since the original DEF model has no potential terms, its background scalar field $\phi_0^*$ is determined by the cosmological evolution. Using \eqref{paras} and \eqref{relation}, we have
\begin{equation}
\xi \simeq  \frac12 \beta_*^2\left(\frac{\phi_0^*}{M_p}\right)^2.
\end{equation}
Applying the constraints in Table \ref{bounds}, we have

\begin{equation}
|\beta_*\frac{\phi_0^*}{M_p}|<2.1\times 10^{-2}.
\end{equation}

In the (original and extended) DEF model, for a sufficiently negative $\beta_*(\lesssim -4)$, the neutron star will cause an activation of the scalar field above its background value, thus influencing its sensitivity. As a result, this  scalarization effect can affect the GW waveforms of  the binary system containing a neutron star \cite{PhysRevD.89.044024,PhysRevD.96.064037,PhysRevD.90.124091}. We ignore the scalarization effect in this paper and leave it in a future work.

\subsection{$f(R)$ gravity}
$f(R)$ gravity is a well studied model to explain the late time accelerated  expansion of the universe \cite{PhysRevD.70.043528}. The action for $f(R)$ gravity takes the form \cite{RevModPhys.82.451}

\begin{equation}
S=\frac{1}{16\pi G_*}\int d^4 x\sqrt{-g }\,f(R)+ S_m[g_{\mu\nu},\Psi_m],
\end{equation}
where $f(R)$ is a function of the Ricci scalar.
After the field redefinition \cite{RevModPhys.82.451}, $f(R)$ gravity can be rewritten as massive Brans-Dicke theory with $\omega(\phi)=0$ and the potential term

\begin{equation}
M(\phi)=\frac{1}{G_*} f(R)-\phi R,
\end{equation}
where the scalar field is defined by $\phi=\frac{1}{G_*}f'(R)$.
Since $\xi=\frac14$ in $f(R)$ gravity, the scalar degree of freedom must be heavy enough  to satisfy the Cassini constraint. Therefore, we should apply the waveforms \eqref{hpfH}
and \eqref{hcfH} to $f(R)$ gravity or use the results of Appendix \ref{two integrals} to obtain the waveforms. Our results are applicable to a general $f(R)$ model. Now, take the $R^2$ model as an example \cite{PhysRevD.97.064016}.

\begin{equation}
f(R) = R+d R^2
\end{equation}
where $d$ is a positive constant.
In the $R^2$ model, the potential term is \cite{RevModPhys.82.451}

\begin{equation}
M(\phi)=-\frac{G_*}{4d}(\phi-\frac{1}{G_*})^2.
\end{equation}
The mass squared of the scalar field is

\begin{equation}
m_s^2=\frac{1}{6d}.
\end{equation}
In the heavy scalar mass situation, the phase of the GW waveforms is complicated and the ppE framework is not applicable. Because of this, there is no available projected GW constraint for $f(R)$ gravity. However, when $m_s>2\pi f$, the phase $\Psi_+$ \eqref{PsiP2} is identical to that for GR except that the chirp mass is multiplied by a factor $(1-\xi)$. Therefore, when

\begin{equation}
m_s>4.1\times 10^{-13}~\text{eV} \left(\frac{f_h}{100\text{Hz}}\right)
\end{equation}
with $f_h$ is the highest sensitive frequency of the GW detector,  $f(R)$ gravity can satisfy the GW constraint.
For the $R^2$ model, we have

\begin{equation}
d<4.2\times 10^{-7}\text{Hz}^{-2}\left(\frac{f_h}{100\text{Hz}}\right)^{-2}.
\end{equation}

\subsection{screened modified gravity}
Screened modified gravity (SMG) is a kind of massive scalar-tensor theories with screening mechanisms to suppress the scalar force in high density regions \cite{Burrage2018}. The action of SMG is given in the Einstein frame \eqref{action}. The behavior of the scalar field in SMG is controlled by the effective potential  \cite{Burrage2018}
\begin{equation}
V_{\text{eff}}(\phi_*)=V(\phi_*)+\rho A(\phi_*),
\end{equation}
where $\rho$ is the conserved density  in the Einstein frame. As a result, the mass and the coupling to matter of the scalar field  can vary in different environments. SMG can behave as a dark energy scalar and avoid solar system constraints. For comparison, the mass of the scalar field in massive Brans-Dicke theory is determined by the bare potential $M(\phi)$ which does not depend on the environment. In this section, we ignore the screening mechanism of the following two SMG models and investigate the consequences. 

\begin{align}
(1):\; V(\phi_*)&=\Lambda^4\exp\left(\frac{\Lambda^{\tilde{\alpha}}}{\phi_*^{\tilde{\alpha}}}\right),& A(\phi_*)&=\exp\left(\frac{\tilde{\beta}\phi_*}{M_p}\right),\\
(2):\; V(\phi_*)&=-\frac12\tilde{\mu}^2\phi_*^2+\frac{\lambda}{4}\phi_*^4,& A(\phi_*)&=1+\frac{\phi_*^2}{2\tilde{M}^2},
\end{align}
Model (1) is the chameleon model \cite{PhysRevLett.93.171104,PhysRevD.69.044026}. $\Lambda$ corresponds to the dark energy scale. $\tilde{\alpha}$ and $\tilde{\beta}$ are the positive dimensionless constants.  Model (2) is the symmetron model \cite{PhysRevLett.104.231301}. $\lambda$ is a positive dimensionless constant. $\tilde{\mu}$ and $\tilde{M}$ are two mass scales.

Using the transformation relation \eqref{relation}, we obtain the potential function $M(\phi)$ and the coupling function $\omega(\phi)$ in the Jordan frame
\begin{align}
(1):\; M(\phi)&=-16\pi\Lambda^4 G_*^2\phi^2\exp\left[\left(\frac{-2\tilde{\beta}\Lambda}{M_p\ln(G_*\phi)}\right)^{\tilde{\alpha}}\right],& \quad \omega(\phi)&=\frac{1}{4\tilde{\beta}^2}-\frac32,\\
(2):\; M(\phi)&=-16\pi G_*^2\phi^2\left[-\frac12\tilde{\mu}^2 \tilde{M}^2(1-G_*\phi)+\frac{\lambda}{4}\tilde{M}^4(1-G_*\phi)^2\right], & \quad \omega(\phi)&=\frac{\tilde{M}^2}{4 M_p^2(1-G_*\phi)}-\frac32  
\end{align}
Since $A(\phi_*)>1$ in these two models, from \eqref{relation}, we have $0<\phi<1/G_*$.
In model (1), $M(\phi)$ is monotonically decreasing. Therefore, we cannot define the mass of the scalar field $\phi$ and our result is not applicable to the chameleon model. This shows that SMG models and massive Brans-Dicke theory have no one-to-one correspondence.

In model (2), from $M'(\phi_0)=0$, we obtain the scalar background

\begin{equation}
G_*\phi_0=1-\frac{\tilde{\mu}^2}{\lambda\tilde{M}^2}
\end{equation}
and the mass squared of the scalar field is

\begin{equation}
m_s^2=\frac{48\pi\phi_0}{2\omega_0+3}G_*^2\tilde{\mu}^2\tilde{M}^2.
\end{equation}
In  \cite{PhysRevLett.104.231301}, the authors impose two relations between the parameters of the symmetron model.
$\tilde{\mu}^2\tilde{M}^2$ is around the current cosmic density
\begin{equation}
\tilde{\mu}^2\tilde{M}^2\sim H_0^2 M_p^2.
\end{equation}
The strength of the scalar force is comparable to gravity
\begin{equation}
\frac{\tilde{\mu}}{\sqrt{\lambda}\tilde{M}^2}\sim \frac{1}{M_p}.
\end{equation}
Using these two relations, we have
\begin{equation}
\omega_0\sim 1,\quad m_s\sim H_0\sim 10^{-33}~\text{eV}.
\end{equation}
It can be seen that these parameters do not satisfy the GW constraints in Table \ref{bounds}.
However, if we consider the screening mechanism, following the discussion of Section VB in \cite{PhysRevD.98.083023}, the GW constraint in Table \ref{bounds} only imposes a weak bound on $\lambda$,
\begin{equation}
\lambda>4.9\times10^{-110}.
\end{equation}
This result indicates the necessity of applying the screening mechanism.

\section{Conclusion and Discussion }\label{con}
We have calculated the GW waveforms of a compact binary on a quasicircular orbit to quadrupole order in massive Brans-Dicke theory. The massless tensor field induces two tensor GW polarizations $h_+$ and $h_\times$. The massive scalar field induces two scalar GW polarizations $h_b$ and $h_L$. 
Our work, with a general coupling function $\omega(\phi)$, confirmed the waveforms of the tensor polarizations obtained in \cite{PhysRevD.85.122005} that assumed a constant coupling function and additionally found the waveforms of the scalar polarizations which contribute to the signal received by a gravitational wave detector.
These will be useful in the search of the massive scalar field. 

Using the SPA method, we also calculate the Fourier transforms of the two tensor GW polarizations $\tilde{h}_+(f)$ and $\tilde{h}_\times(f)$. The expressions of $\tilde{h}_+(f)$ and $\tilde{h}_\times(f)$ depend on  $\omega_0$, but not on the derivative of the coupling function $\omega(\phi)$. This is because the binary system on a circular orbit has no monopole radiation, while the derivative of the coupling function only appears in  monopole terms.  
In the light scalar mass situation the waveforms $\tilde{h}_+(f)$ and $\tilde{h}_\times(f)$ can be mapped to the ppE framework and we obtain the ppE parameters. However, when the scalar mass is comparable to the GW frequency, the phases of these waveforms contain hypergeometric functions which cannot be described by the ppE framework.   Yunes and Pretorius \cite{PhysRevD.80.122003} design this framework, trying to incorporate all metric theories of gravity. We demonstrate explicitly that the applicability of the ppE framework depends on the parameter of the gravitational theory. In the limit $m_s\to \infty$, the binary system has no scalar radiation and the waveforms of $\tilde{h}_+(f)$ and $\tilde{h}_\times(f)$ will  be identical to  that of GR except that the chirp mass $M_c$ is replaced by $M_c(1-\xi)$. Since a binary black hole system also has no scalar radiation, the waveforms of the binary black hole system take the same form. In the limit $m_s=0$, the GW waveforms and radiation power is consistent with those of Brans-Dicke theory. Considering the
potential observations of  the GWs emitted by a black hole-neutron star binary by the future ground based GW detectors, we obtain the constraints on the parameters $\xi$ and $m_s$. Then, we apply our results to specific models, including the DEF model, $f(R)$ gravity, and screened modified gravity. The parameter constraints are based on the previous work \cite{PhysRevD.96.084039} which considers only the phase correction to the tensor polarizations. Therefore, we need to further study the influence of  the amplitude correction and the scalar polarizations on the parameter constraints, especially when the scalar field is heavy. We leave it as a future work.

It can be seen that the sensitivity $s$ of a compact body always appears in the combination of  $(1-2s)$. This is because the source terms in the scalar field equation \eqref{seq} have the form $T-2\phi\frac{\partial T}{\partial \phi}$. The trace of the energy-momentum tensor depends on the scalar field through the mass of the compact body, $T\sim m(\phi)$. Therefore, $T-2\phi\frac{\partial T}{\partial \phi}\sim m(\phi)(1-2s)$. Actually, $(1-2s)$ is proportional to the scalar charge defined by Damour and Esposito-Far\`{e}se in \cite{PhysRevLett.70.2220}. For black holes, a sensitivity of $\frac12$ is equivalent to saying that black holes have no scalar hair \cite{PhysRevLett.108.081103}. However, when the scalar background $\phi_0$ is time dependent \cite{PhysRevLett.83.2699,Horbatsch_2012} or has a  spatial gradient \cite{PhysRevD.87.124020}, a scalar hair can arise \cite{Berti_2015}.
We need to further investigate gravitational waveforms emitted by the hairy black holes in the future work.

\begin{acknowledgments}
T. L. is supported by the National Natural Science Foundation of China (NSFC) Grant No. 12003008 and the China Postdoctoral Science Foundation Grant No. 2020M682393. 
W. Z. is supported by the NSFC  Grants No. 11773028, No. 11633001, No.
11653002, No. 11421303, the Fundamental Research
Funds for the Central Universities and the Strategic Priority
Research Program of the Chinese Academy of Sciences
Grant No. XDB23010200. Y. W. gratefully acknowledges support from the NSFC under Grants No. 11973024 and No. 11690021, and Guangdong Major Project of Basic and Applied Basic Research (Grant No. 2019B030302001). The authors thank the anonymous referees for helpful comments and suggestions.

\end{acknowledgments}

\appendix
\section{Integrals in $h_b$ and $h_L$}\label{int_hbhl}
We follow the method described in Appendix B of \cite{PhysRevD.85.064041} to evaluate the integrals with the Bessel function in \eqref{hb} and \eqref{hl}:
\begin{align}
\begin{split}
I_1=&\int_0^\infty dz J_1(z) \frac{1}{u^2}\omega(t-Ru)^{\frac13}\cos\left(\Phi(t-Ru)\right),\\
I_2=&\int_0^\infty dz J_1(z) \frac{1}{u^3}\omega(t-Ru)^{\frac23}\cos\left(2\Phi(t-Ru)\right),\\
I_3=&\int_0^\infty dz J_1(z) \left(\frac{1}{u^2}-1\right)\frac{1}{u^2}\omega(t-Ru)^{\frac13}\cos\left(\Phi(t-Ru)\right),\\
I_4=&\int_0^\infty dz J_1(z) \left(\frac{1}{u^2}-1\right)\frac{1}{u^3}\omega(t-Ru)^{\frac23}\cos\left(2\Phi(t-Ru)\right),
\end{split}
\end{align}
with $u=\sqrt{1+\big(\frac{z}{m_s R}\big)^2}$ and $\omega(t)=d\Phi(t)/dt$, which cannot be evaluated exactly. Now we will calculate the asymptotic behavior of these integrals in the wave zone $R\to\infty$.

Choosing a parameter $\lambda$ such that $m_s R \lambda\gg 1$ while $\omega R\lambda^2\ll 1$ and splitting $I_3$ into two parts, the first part is
\begin{align}
\begin{split}
&\int_0^{m_s R\lambda}dz J_1(z) \left(\frac{1}{u^2}-1\right)\frac{1}{u^2}\omega(t-Ru)^{\frac13}\cos\left(\Phi(t-Ru)\right)\\
=&-J_0(z)\left(\frac{1}{u^2}-1\right)\frac{1}{u^2}\omega(t-Ru)^{\frac13}\cos\left(\Phi(t-Ru)\right)\Big|_0^{m_sR\lambda}+\cdots\\
=&J_0(m_s R\lambda)\frac{\lambda^2}{(1+\lambda^2)^2}\omega(t-R\sqrt{1+\lambda^2})^{\frac13}\cos(\Phi(t-R\sqrt{1+\lambda^2}))+\cdots
\end{split}
\end{align}
All terms are dependent on $\lambda$. They can be exactly canceled by the second part when we perform integration by parts. Therefore, the asymptotic behavior of $I_3$ is determined by the second part.

Substituting the asymptotic expression of the Bessel function
\begin{equation}
J_\nu(x)\simeq\sqrt{\frac{2}{\pi x}}\cos\left(x-\frac{\nu\pi}{2}-\frac{\pi}{4}\right),
\end{equation}
into the second part, the integral can be approximated by
\begin{align}
\begin{split}\label{i3}
I_3'&=-m_s R\int^\infty_{\sqrt{1+\lambda^2}} du \sqrt{\frac{2}{\pi}}\frac{\cos(m_sR\sqrt{u^2-1}-\frac34\pi)}{\sqrt{m_sR}(u^2-1)^{\frac14}}\frac{\sqrt{u^2-1}}{u^3}\cos(\Phi(t-Ru))\omega(t-Ru)^{\frac13}\\
&=-\frac12\sqrt{\frac{2m_sR}{\pi}}\int^\infty_{\sqrt{1+\lambda^2}} du\frac{(u^2-1)^{\frac14}}{u^3}\omega(t-Ru)^{\frac13}\Re \left[e^{i(m_sR\sqrt{u^2-1}-\frac34\pi+\Phi)}+e^{i(m_sR\sqrt{u^2-1}-\frac34\pi-\Phi)}\right]
\end{split}
\end{align}
where $\Re$ denotes the real part of the argument.
In the previous work \cite{PhysRevD.98.083023}, we have worked out the asymptotic behavior of $I_3'$ when $\omega>m_s$. Now we will focus on the situation $\omega<m_s$. In this situation, the first derivative of the exponential part of the two terms of the integrand cannot vanish on the real axis, we must consider the analytic properties of the exponential part. We will use the method of steepest descent \cite{bender1999advanced}.

The saddle point of the first term $b_1$ is determined by
\begin{equation}
\rho'(b_1)=i(m_sR\frac{b_1}{\sqrt{b_1^2-1}}-\omega(t-Rb_1)R)=0,
\end{equation} 
that is 
\begin{equation}
b_1=\frac{\omega}{m_s^2-\omega^2}e^{i\frac{\pi}{2}},
\end{equation}
where $\rho(u)=i(m_sR\sqrt{u^2-1}-\frac34\pi+\Phi(t-Ru))$ is the exponential part of the first term of \eqref{i3}. Deforming the integration contour to pass this saddle point, we obtain the dominant contribution to the integral $I_3'$ 
\begin{equation}
I_3'\sim \Re\left[-\frac{i}{2}\sqrt{\frac{2m_sR}{\pi}}\frac{(b_1^2-1)^{\frac14}}{b_1^3}\omega(t-Rb_1)^{\frac13}e^{\rho(b_1)}\sqrt{\frac{2\pi}{\rho''(b_1)}}\right]
\end{equation}
\begin{align}
\begin{split}
I_3'&\sim \Re\left[-\frac{i}{2}\sqrt{\frac{2m_sR}{\pi}}\frac{(b_1^2-1)^{\frac14}}{b_1^3}\omega(t-Rb_1)^{\frac13}e^{\rho(b_1)}\sqrt{\frac{2\pi}{\rho''(b_1)}}\right]\\
&=m_s^2\sqrt{m_s^2-\omega(t-R)^2}\omega(t-R)^{-\frac{8}{3}}e^{-R\sqrt{m_s^2-\omega(t-R)^2}}\cos(\Phi(t-R)+\omega R-\frac{\pi}{2}).
\end{split}
\end{align}
Therefore,
\begin{equation}
I_3\sim m_s^2\sqrt{m_s^2-\omega(t-R)^2}\omega(t-R)^{-\frac{8}{3}}e^{-R\sqrt{m_s^2-\omega(t-R)^2}}\cos(\Phi(t-R)+\omega R-\frac{\pi}{2}) .
\end{equation}
In the same way, we can work out the asymptotic behavior of $I_1$,$I_2$ and $I_3$ in the situation $\omega<m_s$
\begin{align}
\begin{split}
I_1\sim&\omega(t-R)\cos(\Phi(t-R))+\omega(t-R)^{\frac13}\sqrt{\frac{m_s^2}{\omega(t-R)^2}-1}~e^{-R\sqrt{m_s^2-\omega(t-R)^2}}\cos(\Phi(t-R)+\omega R+\frac{\pi}{2}),\\
I_2\sim&\omega(t-R)^\frac{2}{3}\cos(2\Phi(t-R))-\omega(t-R)^{\frac{2}{3}}\left(1-\frac{m_s^2}{4\omega(t-R)^2}\right)e^{-R\sqrt{m_s^2-4\omega(t-R)^2}}\cos(2\Phi(t-R)+2\omega R),\\
I_4\sim&-\frac{m_s^2}{4\omega(t-R)^2}\left(1-\frac{m_s^2}{4\omega(t-R)^2}\right)\omega(t-R)^{\frac23}e^{-R\sqrt{m_s^2-4\omega(t-R)^2}}\cos(2\Phi(t-R)+2\omega R).
\end{split}
\end{align}
It can be seen that all  four of these integrals include terms of order $O(e^{-R})$. When substituted into $h_b$ and $h_L$, these terms can be discarded.
In the above calculations, we assumed that $\omega(t-Ru)$ is real. Actually, the imaginary part of $\omega(t-Ru)$ cannot be ignored for some values of $t$. However, in these situations, $e^{\rho(u)}$ in \eqref{i3} will always contribute a term of order $O(e^{-R})$. Therefore, this assumption will not influence the expressions of  $h_b$ and $h_L$.

We collect the asymptotic behavior of these integrals in the situation $\omega>m_s$ to facilitate reference \cite{PhysRevD.98.083023}.
\begin{align}
\begin{split}
I_1 &\sim \omega(t-R)^\frac13\cos(\Phi(t-R))-\omega(t-Ru_1)^{-\frac23}\sqrt{\omega(t-Ru_1)^{2}-m_s^2}\cos\left(\frac{m_s^2R}{\sqrt{\omega(t-Ru_1)^{2}-m_s^2}}+\Phi(t-Ru_1)\right),\\
I_2 &\sim \omega(t-R)^\frac23\cos(2\Phi(t-R))-\omega(t-Ru_2)^{\frac23}\left(1-\frac{m_s^2}{4\omega(t-Ru_2)^2}\right)\cos\Bigg(\frac{m_s^2R}{\sqrt{4\omega(t-Ru_2)^{2}-m_s^2}}+2\Phi(t-Ru_2)\Bigg),\\
I_3 &\sim \frac{m_s^2}{\omega^{\frac83}}\sqrt{\omega^2-m_s^2}\cos\left(\frac{m_s^2R}{\sqrt{\omega^2-m_s^2}}+\Phi\right)\Bigg|_{t-Ru_1},\\
I_4 &\sim \frac{m_s^2}{4\omega^2}\left(1-\frac{m_s^2}{4\omega^2}\right)\omega^\frac23\cos\left(\frac{m_s^2R}{\sqrt{\omega^2-m_s^2}}+\Phi\right)\Bigg|_{t-Ru_2},\\
\end{split}
\end{align}
where $u_n$ is given by
\begin{equation}
u_n=\frac{n\omega(t-R)}{\sqrt{n^2\omega(t-R)^2-m_s^2}}.
\end{equation}

\section{Two integrals}\label{two integrals}
The results of the two integrals in \eqref{hmphase} which are obtained by the software \textit{Mathematica}. $_2F_1(a,b;c;z)$ is the ordinary hypergeometric function.
\begin{align}
\begin{split}
&\int_{m_s}^{\pi f}d\omega~(2\pi f-2\omega)\omega^{-\frac{13}{3}}(1-\frac{m_s^2}{\omega^2})^{\frac32}\\
=& \left[1280 \pi ^3 f^3 \sqrt{\pi ^2 f^2-m_s^2}  \left(176 \pi ^6 f^6-1248 m_s^2 \pi ^4 f^4-1326 m_s^4 \pi ^2 f^2+1183 m_s^6\right) \, _2F_1\left(-\frac{1}{2},-\frac{1}{3};\frac{2}{3};\frac{m_s^2}{\pi ^2 f^2}\right) \right.\\
&+77805 \pi ^3 f^3\sqrt{\pi ^2 f^2-m_s^2}  \left(3 \pi ^6 f^6-100 m_s^2 \pi ^4 f^4-16 m_s^4 \pi ^2 f^2+32 m_s^6\right) \, _2F_1\left(-\frac{5}{6},\frac{1}{2};\frac{1}{6};\frac{m_s^2}{\pi ^2 f^2}\right) \\
&-2240 m_s^2 \pi f \sqrt{\pi ^2 f^2-m_s^2} \left(704 \pi ^6 f^6-2088 m_s^2 \pi ^4 f^4-1014 m_s^4 \pi ^2 f^2+1183 m_s^6\right) \, _2F_1\left(-\frac{1}{2},\frac{2}{3};\frac{5}{3};\frac{m_s^2}{\pi ^2 f^2}\right) \\
&+13832 m_s^2 \pi f\sqrt{\pi ^2 f^2-m_s^2}  \left(209 \pi ^6 f^6-2124 m_s^2 \pi ^4 f^4+300 m_s^4 \pi ^2 f^2+400 m_s^6\right) \, _2F_1\left(-\frac{1}{2},\frac{1}{6};\frac{7}{6};\frac{m_s^2}{\pi ^2 f^2}\right) \\
&+320 \left(\pi ^2 f^2-m_s^2\right) \left(5187 m_s^{2/3}\pi ^{41/6} f^{19/3}  \sqrt{\pi ^2 f^2-m_s^2} \Gamma \left(\frac{5}{3}\right)/\Gamma \left(\frac{25}{6}\right)\right.\\
&+544 \pi ^8f^8 +2840 m_s^2\pi ^6 f^6  -4116  m_s^4\pi ^4 f^4 -1816 m_s^6\pi ^2 f^2  +2548  m_s^8\\
&+39  m_s^2 \pi f \sqrt{\pi ^2 f^2-m_s^2}  \left(32 \pi ^4 f^4-52 m_s^2 \pi ^2 f^2-7 m_s^4\right)  \, _2F_1\left(\frac{1}{2},\frac{2}{3};\frac{5}{3};\frac{m_s^2}{\pi ^2 f^2}\right) \\
&\left.-156 \pi ^3f^3  \sqrt{\pi ^2 f^2-m_s^2} \left(8 \pi ^4 f^4-28 m_s^2 \pi ^2 f^2-7 m_s^4\right)  \, _2F_1\left(-\frac{1}{3},\frac{1}{2};\frac{2}{3};\frac{m_s^2}{\pi ^2 f^2}\right)\right)\\
&-1729 \left(\pi ^2 f^2-m_s^2\right) \left(960m_s^{5/3}\pi ^{35/6} f^{16/3}  \sqrt{\pi ^2 f^2-m_s^2} \Gamma \left(\frac{7}{6}\right)/\Gamma \left(\frac{11}{3}\right) \right.\\
&+135 \pi ^8f^8 -3758 m_s^2\pi ^6 f^6  -6204  m_s^4\pi ^4 f^4 +3432 m_s^6\pi ^2 f^2  +320  m_s^8\\
&\left.\left.+60 m_s^2\pi f \sqrt{\pi ^2 f^2-m_s^2}   \left(11 \pi ^4 f^4-46 m_s^2 \pi ^2 f^2+8 m_s^4\right)  \, _2F_1\left(\frac{1}{6},\frac{1}{2};\frac{7}{6};\frac{m_s^2}{\pi ^2 f^2}\right) \right)\right]\\
&\times\frac{1}{2213120 ~m_s^4 ~\pi ^{16/3}f^{16/3}  \left(\pi ^2 f^2-m_s^2\right)^{3/2}}.
\end{split}
\end{align}

\begin{align}
\begin{split}
&\int_{\frac{m_s}{2}}^{\pi f}d\omega~(2\pi f-2\omega)\omega^{-\frac{11}{3}}(1-\frac{m_s^2}{4\omega^2})^{\frac52}\\
=&\left[- \left(25 m_s^3 \pi  f  \sqrt{4 \pi ^2 f^2-m_s^2}  \left(498688 \pi ^8 f^8-761344 m_s^2\pi ^6 f^6 +98736 m_s^4 \pi ^4 f^4+8756 m_s^6\pi ^2 f^2 -2057 m_s^8\right) \, _2F_1\left(-\frac{1}{2},\frac{1}{3};\frac{4}{3};\frac{m_s^2}{4 \pi ^2 f^2 }\right)\right.\right.\\
&+288  m_s\pi ^3 f^3 \sqrt{4 \pi ^2 f^2-m_s^2} \left(23040 \pi ^8 f^8-153472 m_s^2\pi ^6 f^6 -3564 m_s^4\pi ^4 f^4 +7788 m_s^6\pi ^2 f^2 -935 m_s^8\right)  \, _2F_1\left(-\frac{2}{3},\frac{1}{2};\frac{1}{3};\frac{m_s^2}{4 \pi ^2 f^2 }\right)\\
&+m_s \left(4 \pi ^2 f^2-m_s^2\right) \left(-3317760 \pi ^{10} f^{10} +18984448 m_s^2\pi ^8 f^8  +5410496 m_s^4\pi ^6 f^6  -2107632 m_s^6 \pi ^4 f^4  +177980 m_s^8\pi ^2 f^2+4675 m_s^{10} \right. \\
&-33 m_s^2\pi  f   \sqrt{4 \pi ^2 f^2-m_s^2} \left(40960 \pi ^6 f^6-32064 m_s^2\pi ^4 f^4 +4176 m_s^4\pi ^2 f^2 -85 m_s^6\right) \, _2F_1\left(\frac{1}{3},\frac{1}{2};\frac{4}{3};\frac{m_s^2}{4 \pi ^2 f^2 }\right)\\
&\left.\left. -5161200\ 2^{2/3}  m_s^{4/3}\pi ^{49/6} f^{23/3}\sqrt{4 \pi ^2 f^2-m_s^2} \Gamma \left(\frac{4}{3}\right) /\Gamma \left(\frac{29}{6}\right)\right)\right)\times \frac{35}{m_s^5\pi ^{20/3}f^{20/3}   \left(4 \pi ^2 f^2-m_s^2\right)^{3/2}} \\
&-12903   \left(-\frac{8 \left(367 \pi ^4 f^4-86 m_s^2 \pi ^2 f^2 +7 m_s^4\right) \, _2F_1\left(-\frac{1}{2},\frac{5}{6};\frac{11}{6};\frac{m_s^2}{4 \pi ^2 f^2 }\right)}{\pi ^{17/3}f^{17/3}}+\frac{7000\ 2^{2/3} \pi ^{1/2} \Gamma \left(\frac{5}{6}\right)}{m_s^{5/3} \Gamma \left(\frac{13}{3}\right)}\right.\\
&\left.\left.+\frac{3 \left(232 \pi ^4 f^4-86 m_s^2\pi ^2 f^2 +7 m_s^4\right) \, _2F_1\left(\frac{1}{2},\frac{5}{6};\frac{11}{6};\frac{m_s^2}{4 \pi ^2 f^2 }\right)}{\pi ^{17/3}f^{17/3}}\right)\right]\times\frac{1}{24085600}.
\end{split}
\end{align}


%

\end{document}